\documentclass[preprint,nohyper]{JHEP3} %
\usepackage{epsfig,multicol}
\usepackage{latexsym}

\usepackage{amssymb,amsmath}

\usepackage{multirow,slashbox}
\usepackage{graphics,subfigure}
\usepackage{amsfonts}
\usepackage{multirow}

\newcommand{\dis}[1]{\begin{equation}\begin{split}#1\end{split}\end{equation}}
\newcommand{\beqa}[1]{\begin{eqnarray}#1\end{eqnarray}}
\newcommand{\be}{\begin{equation}}
\newcommand{\ee}{\end{equation}}

\newcommand{\Mp}{M_P}

\newcommand{\D}{\Delta}

\newcommand{\p}{\partial}
\newcommand{\vp}{\varphi}

\newcommand{\calA}{{\cal A}}
\newcommand{\calB}{{\cal B}}

\newcommand{\calP}{{\cal P}}
\newcommand{\calC}{{\cal C}}
\newcommand{\calM}{{\cal M}}

\newcommand{\calG}{{\cal G}}

\newcommand{\xisqggg}{\xi^2_{\sigma\sigma\sigma}}
\newcommand{\xisqggs}{\xi^2_{\sigma\sigma s}}
\newcommand{\xisqgss}{\xi^2_{\sigma s s}}
\newcommand{\xisqsss}{\xi^2_{sss}}
\newcommand{\etagg}{\eta_{\sigma\sigma}}
\newcommand{\etags}{\eta_{\sigma s}}
\newcommand{\etass}{\eta_{s s}}
\newcommand{\eb}{\epsilon_b}
\newcommand{\ebe}{\epsilon_b\epsilon}
\newcommand{\rtebe}{\textrm{sign}(b_\vp)\textrm{sign}\left(\frac{V_\sigma}{V}\right)\sqrt{\epsilon_b \epsilon}}

\newcommand{\sinD}{\sin\Delta}

\newcommand{\cosD}{\cos\Delta}

\newcommand{\tanD}{\tan\Delta}
\newcommand{\tanDsq}{\tan^2\Delta}

\newcommand{\runnzeta}{\frac{d n_\zeta}{d \ln k}}
\newcommand{\runnS}{\frac{d n_S}{d \ln k}}
\newcommand{\runnC}{\frac{d n_C}{d \ln k}}

\newcommand{\epstar}{\epsilon_\vp^*}
\newcommand{\ep}{\epsilon_\vp}
\newcommand{\epe}{\epsilon_\vp^e}
\newcommand{\ecstar}{\epsilon_\chi^*}
\newcommand{\ec}{\epsilon_\chi}
\newcommand{\ece}{\epsilon_\chi^e}

\newcommand{\ete}{\epsilon^e}

\title{Spectral Running and Non-Gaussianity from Slow-Roll Inflation in Generalised Two--Field Models}
\author{Ki-Young Choi \\Department of Physics and Astronomy \\ University of Sheffield \\Sheffield, S3 7RH. UK \\\email{k.choi@sheffield.ac.uk}}
\author{Lisa M.H. Hall and Carsten van de Bruck \\Department of Applied Mathematics \\ University of Sheffield \\Sheffield, S3 7RH. UK \\\email{lisa.hall@sheffield.ac.uk}\\\email{c.vandebruck@sheffield.ac.uk}}

\abstract{Theories beyond the standard model such as string theory motivate low energy effective 
field theories with several scalar fields which are not only coupled through a potential but also 
through their kinetic terms. For such theories we derive the general formulae for the running of 
the spectral indices for the adiabatic, isocurvature and correlation spectra in the case of two 
field inflation. We also compute the expected non-Gaussianity in such models for specific forms 
of the potentials.
We find that the coupling has little impact on the level of non-Gaussianity during inflation.}

\preprint{}
\begin{document}

\section{Introduction}
String theory and theories with supersymmetry usually contain many scalar fields which can 
play an important role in the early Universe. For example, in string theory they often describe the dynamics 
of extra spatial dimensions and other degrees of freedom living in higher dimensions. These 
scalar degrees of freedom couple to matter fields propagating in the three large dimensions we 
perceive, see for example \cite{Lukas}-\cite{Moduli2}. If these extra dimensions exist, they will have had 
an influence on the evolution of the universe at some point. It is usually thought that in particular 
the dynamics of the very early universe is affected by the existence of extra dimensions. As such, they 
will alter the predictions of inflationary cosmology. If apart from the inflaton field(s) other scalar 
fields are present, they will generally alter the evolution of the field(s) driving inflation 
and affect the production of cosmological perturbations. 

Generally speaking, the existence of multiple fields during the inflationary epoch modifys some of 
the single field predictions. For example, apart from the usual adiabatic perturbations produced 
during inflation, there could be isocurvature (entropy) perturbations produced, whose existence 
is constrained by observations (see, for example \cite{Wands02}-\cite{Byrnes:2006fr} and references 
therein). 

Related to this, in single field inflation, the curvature perturbation on constant 
energy density hypersurfaces, $\zeta$, is constant on super-horizon scales and can be 
evaluated at the horizon crossing. However, in the presence of multiple fields during 
the inflationary epoch, $\zeta$ does not remain constant and varies since the non-vanishing 
isocurvature perturbations act as a source term for the change of $\zeta$~\cite{Bellio-Wands95,Bellio-Wands96}.

In order to distinguish between inflationary models, cosmologists need to extract as much 
information as possible from the data. Usually the spectral index and the scalar--tensor ratio 
are used to distinguish between inflationary models. With the advent of precision data (in particular 
the data obtained by WMAP\cite{Spergel:2006hy}), the running of the spectral index can be added as an additional quantity 
in order to distinguish between the models. One can expect considerable improvement with future data 
coming from the PLANCK satellite and the mapping of the large scale structures in the universe, as well 
as a better understanding of small scale clustering as obtained from the Ly$\alpha$ forest. 

Another potential observable which could distinguish between inflationary models is the amount of 
non-Gaussianity generated during inflation (see \cite{non-gaussian1} and references therein). Although 
it is small for many inflationary models (e.g. \cite{Maldacena02}-\cite{Battefeld:2006sz}), there are examples in which 
perturbations show a considerable amount of non-Gaussianity \cite{easther}-\cite{cline}. 

In this paper we study the slow-roll regime of generalised two-field inflation, 
where the scalar fields are also coupled through kinetic terms~\cite{Bellio-Wands95,Bellio-Wands96,Starobinsky01}. 
In general, a multi-field action can be given by
\begin{equation}
  S = \int d^{4}x \, \sqrt{-g} \, \left[ \Mp^2\frac{R}{2} -\frac12
  \calG_{IJ} \, \partial_{\mu} \phi^I \partial^{\mu} \phi^J - W(\phi) \right] .
\end{equation}
For simplicity we consider a system of two scalar fields, whose dynamics are governed by the following action:
\dis{
S=\int d^4 x \sqrt{-g}\left[\Mp^2\frac{R}{2}-\frac12g^{\mu\nu}\partial_\mu\vp\partial_\nu\vp
-\frac12e^{2b(\vp)}g^{\mu\nu}\partial_{\mu}\chi\partial_\nu\chi-W(\vp,\chi) \right]. \label{action}}
In this expression $\Mp=1/\sqrt{8\pi G}$ is the reduced Planck mass.  Note that, in this case, $\calG_{IJ}$ is symmetric.

We obtain the general formulae for the running of the spectral indices (adiabatic, non-adiabatic 
and correlated), following~\cite{DiMarcoFinelli03} and~\cite{DiMarcoFinelli05}. The running is affected by the presence of 
the coupling and therefore future experiments constraining the running of the spectral index will 
give vital information about the existence of fields which couple non-trivially to the field(s) 
driving inflation. Furthermore, we derive the general formulae for non-Gaussianity and apply them to some specific models. 

The paper is organised as follows: In the next Section we summarise the equations governing the background 
and perturbation evolution. 
In Section~\ref{SecRun} we derive the general expressions for the running of the spectral 
indices (for the adiabatic and isocurvature power spectra as well as the correlation spectrum). 
In Section~\ref{SecNonG} we derive the expected non--Gaussianity for two cases. 
In Sections~\ref{SecTheory1} and 
\ref{SecTheory2} we then apply our results to specific examples. Numerical results are presented in Section~\ref{secNumerical}. 
Our conclusions can be found in Section~\ref{SecConc}. Useful formulae are collected in the Appendices.

\section{Background and Perturbation Equations}
In this section we set our notation and briefly review some results found in previous work. It will closely follow \cite{DiMarcoFinelli03} and \cite{DiMarcoFinelli05}.
The equations of motion for the two scalar fields in a Friedmann--Robertson--Walker spacetime 
follow from the action (\ref{action}) and are given by
\dis{
\ddot{\vp}+3H\dot{\vp}+W_\vp=b_\vp e^{2b}\dot{\chi}^2,\\
\ddot{\chi}+(3H+2b_\vp\dot{\vp})\dot{\chi}+e^{-2b}W_\chi=0,
\label{background}
}
where $b_\vp=\frac{\partial b(\vp)}{\partial \vp}$, $W_\vp = \frac{\partial W(\vp,\chi)}{\partial \vp}$, etc. 
Einstein's equations lead to 
\dis{
H^2=\frac{1}{3\Mp^2}\left[\frac12\dot{\vp}^2+\frac12e^{2b}\dot{\chi}^2+W\right], \\
\dot{H}=\frac{1}{2\Mp^2}\left[\dot{\vp}^2+e^{2b}\dot{\chi}^2\right]=-\frac{\dot{\sigma}^2}{2\Mp^2}.
\label{Friedmann}
}
We define e-folding number, $N$, as
\be
N(t_e, t_*) \equiv \int_*^e H dt \label{N}.
\ee

It is useful to separate the perturbations into components of adiabatic and 
isocurvature modes.
Following~\cite{DiMarcoFinelli03,Gordon01}, we define the average (adiabatic) 
and orthogonal (entropy) fields $\sigma$ and $s$ as 
\dis{
d\sigma=&\cos\theta d\vp +\sin\theta e^b d \chi,\\
ds=&e^b\cos\theta d\chi -\sin\theta d \vp,
}
with
\dis{
\cos\theta=\frac{\dot{\vp}}{\sqrt{\dot{\vp}^2+e^{2b}\dot{\chi}^2}},\qquad
\sin\theta=\frac{e^b\dot{\chi}}{\sqrt{\dot{\vp}^2+e^{2b}\dot{\chi}^2}}.
}
These fields satisfy the equations of motion
\dis{
\ddot{\sigma}+3H\dot{\sigma}+W_\sigma=0,\\
\dot{\theta}=-\frac{W_s}{\dot{\sigma}}-b_\vp \dot{\sigma}\sin\theta,
}
where
\dis{
W_\sigma=W_\vp\cos\theta+e^{-b}W_\chi\sin\theta,\\
W_s=-W_\vp\sin\theta+e^{-b}W_\chi\cos\theta.
}

We now turn to cosmological perturbations. We work in the longitudinal gauge, in which the perturbed 
metric has the form
\dis{
ds^2=-(1+2\Phi)dt^2+a^2(1-2\Phi)d{\bf x}^2,
}
where $a=a(t)$ is the scale factor and $\Phi=\Phi(t,{\bf x})$ is the metric perturbation. 
We now calculate the perturbations in the adiabatic and entropy fields.
Instead of working with the field perturbation $\delta \sigma$, it is more convenient to work with the 
Sasaki--Mukhanov variable $Q_\sigma=\delta \sigma+\frac{\dot\sigma}{H}\Phi$. The equation of motion for $Q_\sigma$ 
is given by
\dis{
\ddot{Q}_\sigma &+ 3H\dot{Q}_\sigma +\left[\frac{k^2}{a^2}+W_{\sigma\sigma} +\dot{\theta} \frac{W_s}{\dot{\sigma}}-\frac{1}{\Mp^2a^3}\left(\frac{a^3\dot{\sigma}^2}{H} \right)^{\cdot}-b_\vp\dot{\vp}\frac{W_\chi e^{-b}}{\dot{\sigma}}\sin\theta \right]Q_\sigma \\
&= -2\left(\frac{W_s}{\dot{\sigma}}\delta s \right)^{\cdot}+2\left(\frac{W_s}{\dot{\sigma}}+\frac{\dot{H}}{H} \right)\frac{W_s}{\dot{\sigma}}\delta s,
}
whereas for $\delta s$ it is given by 
\dis{
\ddot{\delta s}+3H\dot{\delta s}+ \left[\frac{k^2}{a^2}+W_{ss}+3\dot{\theta}^2-b_{\vp\vp}\dot{\sigma}^2 + b_\vp^2 g(t) + b_\vp f(t) \right]\delta s = -\frac{k^2}{a^2}\frac{\Phi}{2\pi G}\frac{W_s}{\dot{\sigma}^2},
}
with  
\dis{
g(t)&=-\dot{\sigma}^2(1+3\sin^2\theta),\\
f(t)&=W_\vp(1+\sin^2\theta)-4W_s \sin\theta.
}
We have taken the notation of \cite{DiMarcoFinelli05} and use
\begin{eqnarray}
W_{\sigma\sigma}=W_{\vp\vp} \cos^2\theta+V_{\vp\chi} \sin{2\theta}e^{-b}+V_{\chi\chi}\sin^2{\theta} e^{-2b},\\ 
W_{ss}=W_{\vp\vp} \sin^2\theta-V_{\vp\chi} \sin{2\theta}e^{-b}+V_{\chi\chi}\cos^2{\theta} e^{-2b}.
\end{eqnarray}
We define the curvature and isocurvature fluctuations as
\dis{
\zeta=\frac{H}{\dot{\sigma}}Q_\sigma,\qquad S=\frac{H}{\dot{\sigma}}\delta s,
}
then the time derivative of $\zeta$ is related to $S$ by
\dis{
\dot{\zeta}=\frac{H}{\dot{H}}\frac{k^2}{a^2}\Phi-2\frac{W_s}{\dot{\sigma}}S.
}

During inflation, the fields are assumed to follow the slow-roll limit,
\dis{
\dot{\vp}=\dot{\sigma}\cos\theta\simeq -\frac{W_\vp}{3H}, \qquad
\dot{\chi}=\dot{\sigma}\sin\theta e^{-b} \simeq -\frac{W_\chi}{3H}e^{-2b},
\label{slow-roll}
}
\dis{
H^2(\vp,\chi)\simeq \frac{1}{3\Mp^2}W(\vp,\chi).
}
The slow-roll parameters are defined in Appendix \ref{App1} for convenience.

In the slow-roll limit and on large scales, we find the evolutions of
curvature and isocurvature perturbation can be written in terms of slow-roll
parameters,
\dis{
\dot{\zeta}\simeq BHS,\qquad \dot{S}\simeq -\gamma HS,
\label{evol}
}
where
\dis{
B&=-2\etags-{\rm sign}(b_\vp){\rm sign}\left(\frac{W_\chi}{W}\right)\sqrt{\eb\ec}\sin^2\theta,\\
\gamma&=-\etagg+2\epsilon+\etass +\frac12 {\rm sign}(b_\vp){\rm sign}\left(\frac{W_\chi}{W}\right)\sqrt{\eb\ec}\sin\theta\cos\theta\\
&\qquad +\frac12{\rm sign}(b_\vp) {\rm sign}\left(\frac{W_\vp}{W}\right)\sqrt{\eb\ep}(1+\sin^2\theta). \label{Beq}
}
Generally speaking on large scales the evolutions of adiabatic and isocurvature fluctuations
follow the following set of equations 
\dis{
\dot{\zeta}=\alpha(t)H(t)S,\\
\dot{S}=\delta(t)H(t)S.\label{evolution-of-fluctuations}
}
We use the formalism of transfer matrix~\cite{Wands02},
\dis{
\left( \begin{array}{c}
\zeta(t)\\ S(t)\end{array}\right)
= \left(\begin{array}{cc} 1&T_{\zeta S} \\ 0& T_{SS}\end{array}\right)
\left( \begin{array}{c}
\zeta(t_*)\\ S(t_*)\end{array} \right),
}
where
\dis{
T_{SS}(t_*,t)=\exp\left(\int^t_{t_*}\delta(t')H(t')dt' \right),\\
T_{\zeta S}(t_*,t)=\int^t_{t_*}\alpha(t')H(t')T_{SS}(t_*,t')dt'.
\label{transferfunction}
}
Then the power spectra are
\dis{
\calP_\zeta=(1+T_{\zeta S}^2)\left.\calP_\zeta\right|_*=\calP_\zeta|_*(1+\cot^2\Delta),
\label{Pzeta}
}
\dis{
\calP_S=T^2_{SS}\calP_\zeta|_*,
}
\dis{
\calP_{C}=T_{\zeta S}T_{SS}\calP_\zeta|_*,
}
where the cross-correlation angle $\Delta$ is
\dis{
\cos\Delta=\frac{\calP_{C}}{\sqrt{\calP_{\zeta}\calP_S }}.
}
We note that $T_{\zeta S}=\cot\Delta$ and $\Delta$ has range 
$0\le\Delta\le \pi$ giving positive and negative correlation depending
on the sign of $\cos\Delta$. 
The spectral indices are defined as
\dis{
n_X-1\equiv\frac{d \ln \calP_X}{d \ln k},\qquad X=\zeta, S, \calC.
}
The spectral indices are best written in terms of slow-roll parameters.
They have been calculated in \cite{DiMarcoFinelli05} and are given by 
\begin{eqnarray} 
n_\zeta - 1 &=& -6\epsilon + 4\epsilon(\cos\Delta)^2 + 2\eta_{\sigma\sigma}(\sin\Delta)^2 
+ 4 \eta_{\sigma s} \sin\Delta \cos\Delta + 2\eta_{ss}(\cos\Delta)^2 \nonumber \\
&+& 2~\rm{sign}(b_\vp)\rm{sign}\left(\frac{W_\chi}{W}\right)\sqrt{\epsilon_b 
\epsilon_\chi}(\sin\theta)^2 \sin\Delta\cos\Delta \nonumber \\
&+& \rm{sign}(b_\vp)\rm{sign}\left(\frac{W_\vp}{W}\right)\sqrt{\epsilon_b\epsilon_\vp}\left(1
+\sin^2\theta\right)\cos^2\Delta \nonumber \\ &-& \rm{sign}(b_\vp)\rm{sign}\left(\frac{W_\chi}{W}\right)
\sqrt{\epsilon_b \epsilon_\chi}sin\theta\cos\theta\sin^2 \Delta,
\end{eqnarray}
\begin{eqnarray}
n_S - 1 = -2\epsilon + 2 \eta_{ss} 
+ \rm{sign}(b_\vp)\rm{sign}\left(\frac{W_\vp}{W}\right)\sqrt{\epsilon_b \epsilon_\vp}(1+\sin^2 \theta),
\end{eqnarray}
and 
\begin{eqnarray}
%n_C - 1 = n_S - 1 + \left(2\eta_{\sigma s} + \rm{sign}(b_\vp)\rm{sign}\left(\frac{W_\chi}{W}\right)\sqrt{\epsilon_b \epsilon_\chi}\sin^2\theta)\right)\tan\Delta
n_C - 1 = n_S - 1 -B\tan\Delta.
\label{n_C}
\end{eqnarray}
In these expressions, all the slow-roll parameters are evaluated at horizon crossing. 

\section{Running of the Spectral Indices}
\label{SecRun}
To calculate the running of the spectral indices we need to know the 
second-order slow-roll parameters and the derivative of first-order 
slow-roll parameters and the transfer functions.
For convenience, we list them extensively in Appendix \ref{App1}.

When $b_\vp \ne 0$ we have 14 parameters at second-order in general, as listed in Table~\ref{Tab2}.
However, considering the component fields ($\sigma$,$s$), $H$, $T_{SS}$ and $\xi^2_{sss}$
do not appear in the running of spectral indices, which reduces the parameters to 11.  
Therefore the running spectral indices are given by: $\theta$, $\Delta$ and 9 slow-roll parameters. 
We define the running spectral indices $\alpha_X$ as 
\dis{
\alpha_X\equiv \frac{d n_X}{d \ln k},\qquad X=\zeta, S, \calC. 
}

To leading order in slow roll, the runnings of the spectral indices 
after inflation are:
\dis{
\alpha_\zeta\equiv\runnzeta=&\left(\runnzeta\right)_{b=0}
+\frac{1}{16}\Big[5-3\cos2\theta+8(1-\cos2\theta)(\sin2\theta\sin2\D+\cos2\theta\cos2\D)\\
   -2\sin2&\theta(2-3\cos2\theta+\cos^22\theta)\sin4\D-\{3+\cos2\theta(3-6\cos2\theta+2\cos^22\theta)\}\cos4\D   \Big]\ebe\\
&-\frac18\Big[1+\cos2\theta+2\cos2\D+(1-\cos2\theta)(\sin2\theta\sin2\D+\cos2\theta\cos2\D) \Big]\xi_b\epsilon\\
&+\Big[2\cos^3\theta+\cos\theta(3-2\cos2\theta)\cos2\D+\cos\theta(-2+\cos2\theta)\cos4\D\\
&+2\sin^3\theta(2\sin2\D-\sin4\D) \Big]\rtebe\epsilon\\
&+\Big[\Big\{-\cos\theta+\sin\theta(1-\cos2\theta)(-\sin2\D+\frac12\sin4\D)+\frac12\cos\theta(-3+2\cos2\theta)\cos2\D \\
&+\frac12\cos\theta(2-\cos2\theta)\cos4\D \Big\}\etagg
+\Big\{ \frac52\sin\theta+\frac12\cos\theta(3+\cos2\theta)\sin2\D\\
&-\cos\theta(2-\cos2\theta)\sin4\D-\frac12\sin\theta(6+\cos2\theta)\cos2\D+\sin\theta(1-\cos2\theta)\cos4\D  \Big\}\etags\\
&+\Big\{\frac32\cos\theta+\frac12\sin\theta(3+\cos2\theta)\sin2\D-\sin^3\theta\sin4\D+\cos^3\theta\cos2\D\\
&+\frac12\cos\theta(-2+\cos2\theta)\cos4\D  \Big\}\etass\Big]\rtebe,\\
%%%%%%%
\alpha_S\equiv\runnS=
&\left(\runnS\right)_{b=0} +\sin^2\theta \cos^4\theta \epsilon_b\epsilon  -\frac12(1+\sin^2\theta)\cos^2\theta\xi_b \epsilon
+ \left[2\cos\theta \epsilon  -(1+ \sin^2 \theta)\cos\theta\etass\right.\\
&+\left. \sin\theta(-1+3\sin^2\theta) \eta_{\sigma s} 
 +2\,\cos^3\theta\eta_{ss}  \right]\rtebe,\\
%%%%%%
\alpha_C\equiv\runnC=
&\runnS+ \tanD\Big[ 4\eta_{\sigma s}(\eta_{\sigma \sigma} - \eta_{ss}) -2\xi_{\sigma\sigma s}^2 + \frac12\sin^2\theta\sin2\theta\cos2\theta\ebe
 -\frac12\sin^3\theta\cos\theta\xi_b\epsilon\\
& +\Big\{\sin\theta(3\cos^2\theta-1)\etass-5\sin^2\theta\cos\theta\etags \Big\} \rtebe\Big]\\
&+\tanDsq\Big[-4\eta_{\sigma s }^2-4\eta_{\sigma s }\sqrt{\epsilon_b\epsilon}\sin^3\theta -\sin^6\theta\epsilon_b\epsilon \Big],}
where 
\dis{
\left(\runnzeta\right)_{b=0}=
&2(-7+4\cos2\D-\cos4\D)\epsilon^2+2(4-3\cos2\D+\cos4\D)\epsilon\etagg
+4(2\sin2\D-\sin4\D)\epsilon\etags\\
&+2(2+\cos2\D-\cos4\D)\epsilon\etass
 +\sin^22\Delta(\etagg^2+\etass^2)+ 4\cos^22\Delta\etags^2-(1-\cos4\D)\etagg\etass   \\
&+2\sin4\D(\etagg\etags-\etags\etass) -(1-\cos2\D)\xi_{\sigma\sigma\sigma}-2\sin2\Delta\xi_{\sigma\sigma s}^2 -(1+\cos2\D)\xi_{\sigma s s}^2,\\
%%%
\left(\runnS\right)_{b=0} =& 
-8 \epsilon^2+4\epsilon(\eta_{\sigma\sigma} +\eta_{ss})+4\eta_{\sigma s}^2 -2\xi_{\sigma ss}^2.
}

All slow-roll parameters are evaluated at horizon crossing.
We note that the running spectral index of isocurvature perturbation, $\alpha_S$, is independent of $\Delta$, which means that the $\alpha_S$ is determined
at horizon crossing and does not change thereafter.

\begin{table}[tb]
\begin{center}
%\begin{tabular}{|c||cc|cc|ccc|cc|cccc|c|}
\begin{tabular}{|c||cccccccccccccc|}
\hline
$\vp$,$\chi$ & \multirow{2}{*}{$H$} & \multirow{2}{*}{$T_{SS}$} & \multirow{2}{*}{$\Delta$} & $\epsilon_\vp$ & $\epsilon_\chi$ & $\eta_{\vp\vp}$ & $\eta_{\vp\chi}$ & $\eta_{\chi\chi}$ & 
\multirow{2}{*}{$\eb$} & \multirow{2}{*}{$\xi_b$} & $\xi^2_{\vp\vp\vp}$ & $\xi^2_{\chi\chi\chi}$ & $\xi^2_{\vp\vp\chi\chi}$ & $\xi^2_{\vp\chi\chi\chi}$ \\
$\sigma$,$s$ & & & & $\epsilon$ & $\theta$ & $\etagg$ & $\etags$ & $\etass$ & 
& & $\xisqggg$ & $\xisqggs$ & $\xisqgss$ & $\xisqsss$ \\
\hline
\end{tabular}
\caption{Full set of model parameters to second order, listed by physical fields ($\vp$,$\chi$) or component fields ($\sigma$,$s$).  
$H$, $T_{SS}$ and $\xisqsss$ are not required for the spectral running and can be disregarded.} 
\label{Tab2} 
\end{center} 
\end{table}

\section{Non-Gaussianity During Inflation}
\label{SecNonG}
Another potential discriminator between inflationary models is the level of 
non--Gaussianity produced during inflation. 
Current observations limit the nonlinearity parameter, $|f_{\rm NL}|<100$~\cite{Creminelli:2006rz}.  A perfect CMB experiment cannot hope to detect $|f_{\rm NL}|<3~$\cite{Komatsu:2001rj}.
Using the $\delta$N formalism \cite{starob85,ss1,Sasaki:1998ug,lms,lr}, the non-linearity parameter is 
given by~\cite{Seery05}
\dis{
-\frac65f_{\rm NL}=\frac{\calA^{IJK}N_{,I} N_{,J} N_{,K}}{( N_{,I} N_{,J}\calG^{IJ})^2\sum_i k_i^3 } 
+\frac{\calG^{IM}\calG^{KN} N_{,I} N_{,K} N_{,MN} }{( N_{,I} N_{,J}\calG^{IJ})^2}\label{fNL},
}
which is valid in the slow-roll limit.
The first term, which we call $-\frac65f_{\rm NL}^{(3)}$, can be written as~\cite{VernizziWands},
\dis{
-\frac65f_{\rm NL}^{(3)}=\frac{r}{16}(1+f),
\label{fNL3}
}
where $r\equiv \calP_T/\calP_\zeta$ is tensor-to-scalar ratio and 
$f$ is a function of the momentum 
triangle with the range of values $0\le f\le\frac56$~\cite{Maldacena02}.
In Appendix \ref{AppC}, we have shown Eqn.(\ref{fNL3}) is valid for non-canonical kinetic terms.
From \cite{DiMarcoFinelli05}, the tensor-to-scalar ratio is calculated to be
$r\lesssim 16\epsilon$
and therefore we can approximate
\be
\left| -\frac65f_{\rm NL}^{(3)} \right| \lesssim\epsilon
\ee
which is certainly too small to be observable in CMB experiments. 
Also see \cite{Zaballa:2006pv}. 
Due to this, we will concentrate on the second term, $f_{\rm NL}^{(4)}$, only in the next sections and consider two cases: separable potentials of product and sum.

\subsection{Product Potential $W(\vp,\chi)=U(\vp)V(\chi)$}
We first consider the case of a separable potential by product, for which 
we derive the analytic formula for the non-linear parameter $f_{\rm NL}$.
To use Eq.~(\ref{fNL}), we need to know the dependence of the number of 
e-foldings on the fields $\vp$ and $\chi$.
Following the calculations in~\cite{Bellio-Wands96}, we can obtain the 
derivatives of $N$ by $\vp_*$ and $\chi_*$.

First we find the number of e--foldings,
\be
N(\vp_*,\chi_*)= -\frac{1}{\Mp^2} \int^e_*  \frac{U}{U_\vp} d\vp
= -\frac{1}{\Mp^2} \int^e_*  \frac{V}{V_\chi} e^{2b(\vp)} d\chi,
 \ee
where superscript (or subscript) $^e$ and $^*$ denotes the values evaluated 
at the end of inflation and horizon crossing respectively.
From the equations of motion, in the slow-roll regime, it is possible to find a constant of motion along the trajectory, as in~\cite{Bellio-Wands96}:
\be
 C_1\equiv -\frac{1}{\Mp^2} \int e^{-2b(\vp)} \frac{U d \vp}{U_\vp} + \frac{1}{\Mp^2} \int
  \frac{V d \chi}{V_\chi}  .
\label{integral1}
\ee

\noindent Using this constant of motion, in the slow-roll regime we find the 
first derivatives of $N(t_e,t_*)$ with respect to the fields,
\dis{
\frac{\p N }{\p\vp_*}&=\frac{{\rm sign}(U/U_\vp)}{\Mp\sqrt{2\epstar}}\left[1-\frac{\ece}{\ete}
e^{2b_e-2b_*} \right],\\
\frac{\p N }{\p\chi_*}&=\frac{{\rm sign}(V/V_\chi)}{\Mp\sqrt{2\ecstar}}\left[\frac{\ece}{\ete}
e^{2b_e-b_*} \right].\label{non-gaussian1}
}

\noindent From the first derivatives we can find the second derivatives,
 \begin{eqnarray}
\Mp^2 \frac{\partial^2 N}{\partial \vp_*^2} &=&
1
 - \frac{\eta^*_{\vp\vp}}{2 \epsilon^*_\vp} +
 \frac12{\rm sign}\left(b_\vp\right) {\rm sign}\left(\frac{U_\vp}{U}\right)\sqrt{\frac{\epsilon^*_b}{\epsilon^*_\vp}}
 \frac{\epsilon^e_\chi}{\epsilon^e} e^{2b_e-2b_*}
 - \frac{1}{\epsilon^*_\vp}e^{4b_e-4b_*} {\cal A}_P
, \nonumber\\
\Mp^2 \frac{\partial^2 N}{\partial \chi_*^2} &=& 
e^{2b_*} \left[
\left(1
 - \frac{\eta^*_{\chi\chi}}{2 \epsilon^*_\chi}\right) 
 \frac{\epsilon^e_\chi}{\epsilon^e} e^{2b_e-2b_*}
 - \frac{1}{\epsilon^*_\chi}e^{4b_e-4b_*} {\cal A}_P
\right]
,\label{non-gaussian2}\\
\Mp^2 \frac{\partial^2 N}{\partial \vp_* \partial \chi_*} &=&
{\rm sign}\left(\frac{U_\vp}{U}\right){\rm sign}\left(\frac{V_\chi}{V}\right)\frac{1}{\sqrt{\epsilon^*_\vp \epsilon^*_\chi}} e^{4b_e-3b_*} {\cal A}_P
,\nonumber
 \end{eqnarray}
where
\begin{eqnarray}
\eta_{ss} &\equiv&
\frac{(\epsilon_\chi\eta_{\vp\vp}+\epsilon_\vp\eta_{\chi\chi})}{\epsilon}, \\
\calA_P&\equiv&-\frac{\epsilon^e_\vp \epsilon^e_\chi}{(\epsilon^e)^2}
\left[
\eta_{ss}^e - 
\frac12{\rm sign}\left(b_\vp\right) {\rm sign}\left(\frac{U_\vp}{U}\right) \frac{(\epsilon^e_\chi)^2}{\epsilon^e} \sqrt{\frac{\epsilon^*_b}{\epsilon^*_\vp}} 
- 4 \frac{\epsilon^e_\vp \epsilon^e_\chi}{\epsilon^e}
\right].
\end{eqnarray}
We have followed the notation of \cite{VernizziWands} as much as possible, for reasons that become apparent in the next section.

From Eqs.~(\ref{non-gaussian1}) and (\ref{non-gaussian2}) 
we find the second term of Eq.~(\ref{fNL}),
\begin{eqnarray}
-\frac{6}{5}  f^{(4)}_{\rm NL}&=& 
\frac{2 e^{-2b_e+2b_*}}{\left( \frac{u^2\alpha^2}{\epsilon_\vp^*}
+ \frac{v^2}{\epsilon_\chi^*} \right)^2}
\left[
\frac{u^3\alpha^3}{\epsilon^*_\vp}
\left(1
 - \frac{\eta^*_{\vp\vp}}{2 \epsilon^*_\vp}
\right)
+ \frac{v^3}{\epsilon^*_\chi}
\left(1
 - \frac{\eta^*_{\chi\chi}}{2 \epsilon^*_\chi }
\right) \right.
\nonumber
\\
&&
\quad\left.
+ \frac12 {\rm sign}\left(b_\vp\right) {\rm sign}\left(\frac{U_\vp}{U}\right)
\frac{vu^2\alpha^2}{(\epsilon_\vp^*)^2} \sqrt{\epsilon^*_b\epsilon^*_\vp}
- \left( \frac{u\alpha}{{\epsilon^*_\vp}}
- \frac{v}{{\epsilon^*_\chi}}
\right)^2
e^{2b_e-2b_*} \calA_P
\right]
,
\label{fNL4}
\end{eqnarray}
with the definitions
\be
u\equiv\frac{\epsilon_\vp^e}{\epsilon^e}, \quad v\equiv\frac{\epsilon_\chi^e}{\epsilon^e},
\quad \alpha\equiv e^{-2b_e+2b_*}\left[1+\frac{\epsilon_\chi^e}{\epsilon_\vp^e}
\left(1-e^{2b_e-2b_*}\right)\right].
\ee
When $b(\vp)=0$, then $\alpha=1$ and $\epsilon_b=0$, which means that the symmetry between $\vp$ and $\chi$ in Eq.(\ref{fNL4}) is restored.

\subsection{Sum Potential $W(\varphi,\chi) = U(\varphi) + V(\chi)$}
The second case we consider consists of separable potential models by sum.  
Similar cases with canonical kinetic terms were previously
studied in \cite{VernizziWands}, where it was shown that they do not generate 
significant non-Gaussianity. We will now derive the general formula 
for $f_{\rm NL}$ in the presence of non-canonical kinetic terms, closely following the 
derivation found in the above-mentioned paper.

The total number of e--foldings along a trajectory in slow-roll regime
is given by
\be
N(\vp_*,\chi_*)=- \frac{1}{\Mp^2} \int^e_*  \frac{U}{U_\vp} d\vp -
\frac{1}{\Mp^2} \int^e_* e^{2b(\vp)} \frac{V}{V_\chi} d\chi.
 \ee
We note that in principle $\vp$ in the integration of $\chi$ can be 
re-written in terms of 
$\vp_*,\ \chi_*$ and $\chi$ along the trajectory using equations of motion.
For sum potentials, $\vp$ and $\chi$ have the relation along the
trajectory through
\dis{
\int^\vp_{\vp_*}\frac{e^{-2b(\vp)}}{U_\vp}d\vp=\int^\chi_{\chi_*}\frac{1}{V_\chi}d\chi. 
}
As before, the integral of motion along the trajectory leads to a constant of motion
\be
 C_2\equiv -\Mp^2 \int e^{-2b(\vp)} \frac{d \vp}{U_\vp} + \Mp^2 \int
  \frac{d \chi}{V_\chi}  .
\label{integral2}
\ee

\noindent This enables to derive the first derivatives of $N$ 
\begin{eqnarray}
\Mp \frac{\partial N}{\partial \vp_*}&=&
 {\rm sign}\left(\frac{U_\vp}{W}\right) \frac{1}{\sqrt{2\epsilon_\vp^*}W^*} \left(U^*+Z^e\right) - {\cal G} , \label{dNdp} \nonumber \\
\Mp \frac{\partial N}{\partial \chi_*}&=&
{\rm sign}\left(\frac{V_\chi}{W}\right)\frac{e^{b_*}}{\sqrt{2\epsilon_\chi^*}W^*}
 \left(V^*- Z^e\right) - {\cal H}, \label{dNdc}
\end{eqnarray}
using the definitions,
\begin{eqnarray}
Z^e &=& \frac{(V^e {\epsilon^e_\vp} - U^e
{\epsilon^e_\chi})}{\epsilon^e} e^{2b_e-2b_*} , \label{Z2}\\
{\cal G}(\vp_*,\chi_*) &\equiv&\frac{1}{\Mp} \int^e_* 2b_\vp e^{2b(\vp)} \frac{V}{V_\chi} \frac{\partial\vp}{\partial\vp_*} d\chi \, ,\label{Geqn}\\
{\cal H}(\vp_*,\chi_*) &\equiv&\frac{1}{\Mp} \int^e_* 2b_\vp e^{2b(\vp)} \frac{V}{V_\chi} \frac{\partial\vp}{\partial\chi_*} d\chi \, . \label{Heqn} 
\end{eqnarray}
In the same way the second derivatives can be derived: 
 \begin{eqnarray}
 \label{d2Ndp2}
\Mp^2 \frac{\partial^2 N}{\partial \vp_*^2} &=&
1
 - \frac{\eta^*_{\vp\vp}}{2 \epsilon^*_\vp} \frac{U^*+Z^e}{W^*} +
 {\rm sign}\left(\frac{U_\vp}{W}\right) \frac{\Mp}{W^*\sqrt{2\epsilon_\vp^*}}
 \frac{\partial Z^e}{ \partial \vp^*} - \Mp\frac{\partial {\cal G}}{\partial \vp_*}
, \nonumber \\
 \label{d2Ndc2}
\Mp^2 \frac{\partial^2 N}{\partial \chi_*^2} &=& e^{2b_*}\left[1 -
\frac{\eta^*_{\chi\chi}}{2 \epsilon^*_\chi} \frac{V^*-Z^e}{W^*} -
 {\rm sign}\left(\frac{V_\chi}{W}\right) \frac{\Mp e^{-b_*}}{W^*\sqrt{2\epsilon_\chi^*}}
 \frac{\partial Z^e}{ \partial \chi^*}\right] - \Mp\frac{\partial {\cal H}}{\partial \chi_*},
\nonumber \\
 \label{d2Ndcdp}
\Mp^2 \frac{\partial^2 N}{\partial \chi_*\partial \vp_*} &=&
 {\rm sign}\left(\frac{U_\vp}{W}\right)\frac{\Mp}{W^*\sqrt{2\epsilon_\vp^*}}
 \frac{\partial Z^e}{ \partial \chi^*} - \Mp\frac{\partial {\cal G}}{\partial \chi_*},
 \\
 \label{d2Ndpdc}
\Mp^2 \frac{\partial^2 N}{\partial \vp_*\partial \chi_*} &=&
e^{b_*}\left[
\frac{1}{2} {\rm sign}(b_\vp) {\rm sign}\left(\frac{V_\chi}{W}\right)\sqrt{\frac{\epsilon_b^*}{\epsilon_\chi^*}}\frac{V^*-Z^e}{W^*} 
- {\rm sign}\left(\frac{V_\chi}{W}\right)\frac{\Mp}{W^*\sqrt{2\epsilon_\chi^*}} \frac{\partial Z^e}{ \partial \vp^*} 
\right]
-\Mp \frac{\partial {\cal H}}{\partial \vp_*}. \nonumber
 \end{eqnarray}
The order of the second derivatives, 
$\frac{\partial^2 N}{\partial \vp_*\partial \chi_*} = \frac{\partial^2 N}{\partial \chi_*\partial \vp_*}$, 
and the last two equations emphasise that ${\cal G}$ and ${\cal H}$ are not independent 
quantities.
In fact, they are related by
\begin{eqnarray}
\Mp \frac{\partial {\cal G}}{\partial \chi_*} &=& \Mp \frac{\partial {\cal H}}{\partial\vp_*}
 - {\rm sign}(b_\vp){\rm sign}\left(\frac{V_\chi}{W}\right)\frac{e^{b_*}}{2}\frac{V^*}{W^*}\sqrt{\frac{\epsilon_b^*}{\epsilon_\chi^*}}.
\end{eqnarray}
A similar expression relating $\frac{\partial {\cal G}}{\partial \vp_*}$ to $\frac{\partial {\cal H}}{\partial \vp_*}$  and $\frac{\partial {\cal H}}{\partial \chi_*}$ can be obtained, but it is irrelevant for our purposes and is therefore not given.
From the definition of $Z^e$ in Eq.~(\ref{Z2}), we can calculate
\begin{eqnarray}
{\rm sign}\left(\frac{U_\vp}{W}\right)
\sqrt{\epsilon_\vp^*}\frac{\partial Z^e}{ \partial \vp_*} 
= \frac{\sqrt2}{\Mp} W^*\left[ \calA_S
+ \calB_S
+ \calC_S \right], \\
{\rm sign}\left(\frac{V_\chi}{W}\right)
\sqrt{\epsilon_\chi^*}\frac{\partial Z^e}{ \partial \chi_*} e^{-b_*} 
= -\frac{\sqrt2}{\Mp} W^*\left[ \calA_S 
+ \calB_S \right],
\end{eqnarray}
where we define
\begin{eqnarray}
\calA_S &\equiv& - \frac{W_e^2}{W_*^2} \frac{\epsilon^e_\vp
\epsilon^e_\chi}{\epsilon^e}
 \left(1 - \frac{\eta_{ss}^e}{\epsilon^e} -\frac12{\rm sign}(b_\vp){\rm sign}\left(\frac{U_\vp}{W}\right)\frac{\epsilon_\chi^e}{(\epsilon^e)^2} \sqrt{\epsilon_b^e \epsilon_\vp^e} \right) e^{4b_e-4b_*} \, , \\
\calB_S &\equiv& \frac12 {\rm sign}(b_\vp){\rm sign}\left(\frac{U_\vp}{W}\right)\frac{\epsilon_\chi^e}{\epsilon^e} \sqrt{\epsilon_b^*\epsilon_\vp^*} \frac{W^e}{W_*^2} Z^e e^{2b_e-2b_*} \, , \\
\calC_S &\equiv& -\frac12 {\rm sign}(b_\vp){\rm sign}\left(\frac{U_\vp}{W}\right) \frac{Z^e}{W^*} \sqrt{\epsilon_b^*\epsilon_\vp^*}.
\end{eqnarray}

From the results of the first and second derivatives, it is straightforward to calculate

\begin{eqnarray}
-\frac{6}{5}  f^{(4)}_{\rm NL}&=& 
\frac{2}{\left( \frac{u^2\alpha_u^2}{\epsilon_\vp^*}
+ \frac{v^2\alpha_v^2}{\epsilon_\chi^*} \right)^2}
\left[
\frac{u^2\alpha_u^2}{\epsilon^*_\vp}
\left(1
 - \frac{\eta^*_{\vp\vp}}{2 \epsilon^*_\vp}
u - \Mp\frac{\partial {\cal G}}{\partial \vp_*} 
+\frac{\calC_S}{\epsilon_\vp^*}
\right)
\right.
\nonumber \\
&& 
%\quad 
\left.
%\right.
%\\
%&& \quad \left.
- 2\Mp {\rm sign}\left(\frac{U_\vp}{W}\right){\rm sign}\left(\frac{V_\chi}{W}\right) \frac{u\alpha_u}{\sqrt{\epsilon_\vp^*}} \frac{v\alpha_v e^{-b_*}}{\sqrt{\epsilon_\chi^*}} \frac{\partial {\cal G}}{\partial \chi_*}
\right.
\\
&& \quad \quad \quad \left.
+ \frac{v^2\alpha_v^2}{\epsilon^*_\chi}
\left(1
 - \frac{\eta^*_{\chi\chi}}{2 \epsilon^*_\chi }
v -\Mp\frac{\partial H}{\partial \chi_*}e^{-2b_*}
\right)
+ \left( \frac{u\alpha_u}{{\epsilon^*_\vp}}
- \frac{v\alpha_v}{{\epsilon^*_\chi}}
\right)^2
\left(\calA_S+\calB_S\right)
\right]
,
\nonumber
\end{eqnarray}
where we have defined
\begin{eqnarray}
u \equiv \frac{U^*+Z^e}{W^*}, \quad \quad
v \equiv \frac{V^*-Z^e}{W^*},
\end{eqnarray}
\begin{eqnarray}
\alpha_u \equiv 1+{\rm sign}\left(\frac{U_\vp}{W}\right)\frac{\sqrt{2\epsilon_\vp^*}}{u} {\cal G} , \quad \quad
\alpha_v \equiv 1+{\rm sign}\left(\frac{U_\chi}{W}\right)\frac{\sqrt{2\epsilon_\chi^*}e^{-b_*}}{v} {\cal H}.
\end{eqnarray}

When $b(\vp)=0$, this calculation of $f^{(4)}_{\rm NL}$ simplifies, since
\begin{eqnarray}
\calB_S=\calC_S={\cal G}={\cal H}=0 , \quad \quad \alpha_u=\alpha_v=1,
\end{eqnarray}
and the result is identical to that found in \cite{VernizziWands}.

\section{Examples of Scalar-Tensor Theories with Product Potentials}
\label{SecTheory1}

At first, we consider a separable potential by product with exponents in the $\vp$ field, 
as in the case of a massless dilaton, $\vp$,
\begin{eqnarray}
W(\varphi,\chi)=U(\vp)V(\chi)=e^{4c(\varphi)}V(\chi)\label{app1}.
\end{eqnarray}
When considering the physical fields ($\vp$,$\chi$), only $H$ and $T_{SS}$ can be discarded.
Therefore, the set of parameters consists of $\Delta$ and 11 slow-roll parameters as shown in Table~\ref{Tab2}.
However, with the potential chosen above, the set reduces to only 5 independent slow-roll parameters, which are useful for the running spectral indices:
\beqa{ 
\epsilon_{\varphi} = 8 M^2_{\rm P} c_{\varphi}^2, && \qquad \epsilon_{\chi} = \frac{ M^2_{\rm P}}{2}\left(\frac{V_{\chi}}{V}\right)^2 e^{-2b}, \label{par_1}
\\
%\be \eta_{\varphi \varphi} = 4 M^2_{\rm P}   c_{\varphi\varphi} + 2 \epsilon_{\varphi} \label{par_3} \, ,\ee
\eta_{\chi \chi} = M^2_{\rm P}\left( \frac{V_{\chi\chi}}{V}\right)
e^{-2b},&&\qquad
\xi_{\chi\chi\chi}^2=M^4_{\rm P}\left( \frac{V_{\chi\chi\chi V_\chi}}{V^2}\right) e^{-4b}, \label{par_4} 
\\
\xi_c&=&8 M_{\rm P}^2 c_{\vp\vp},
%, && \qquad 
%\xi_{cc}=\frac{8}{\sqrt2}M^3_{\rm P}c_{\vp\vp\vp}
}
since the remaining 8 are not independent:
\be \epsilon_c=\epsilon_\varphi\, , \ee  
\dis{
\eta_{\varphi \varphi} = \frac12 \xi_c + 2 \epsilon_{\varphi} \, , \qquad
\eta_{\varphi \chi} = 2  {\rm sign}(c_{\varphi}){\rm sign}
\left(\frac{V_{\chi}}{V}\right) \sqrt{\epsilon_{\varphi}\epsilon_{\chi}}
=\epsilon\sin2\theta,  \label{par_3}
}
\be \xi_{\vp\vp\vp}^2=4\epsilon_\vp^2+3\xi_c\epsilon_\vp+\xi_{cc}\sqrt{\epsilon_\vp}\, , \quad \quad \xi_{\vp\vp\chi\chi}^2=4\epsilon_\vp\epsilon_\chi+\epsilon_\chi\xi_c
\, , \ee

\be 
\xi_{\vp\chi\chi\chi}^2=2{\rm sign}(c_{\varphi}){\rm sign}
\left(\frac{V_{\chi}}{V}\right) \sqrt{\epsilon_{\varphi}\epsilon_{\chi}}\eta_{\chi\chi}\, . \ee

\begin{table}
\begin{center}
\begin{tabular}{|c|cc|cc|}
\hline
Model & 1a &1b & 2a& 2b\\ \hline
$b(\vp)$ & $-\beta\frac{\vp}{\Mp}$ &  $-\beta\frac{\vp}{\Mp}$ & $-\beta\frac{\vp^2}{\Mp^2}$ &  $-\beta\frac{\vp^2}{\Mp^2}$\\
$V(\chi)$ & $\frac{\lambda}{4}\chi^4$ & $\frac12m_\chi^2\chi^2$ &$\frac{\lambda}{4}\chi^4$ & $\frac12m_\chi^2\chi^2$ \\ 
$\eta_{\chi\chi}$ & $\frac32\epsilon_\chi$ & $\epsilon_\chi$ &$\frac32\epsilon_\chi$ & $\epsilon_\chi$ \\
$\xi^2_{\chi\chi\chi} $ & $\frac32\epsilon_\chi^2$ & $0$ &$\frac32\epsilon_\chi^2$ & $0$ \\
$\xi_b$ & $0$ & $0$ &$-16\beta$& $-16\beta$\\ \hline
\end{tabular}
\caption{Four Jordan-Brans-Dicke models considered in the text 
and their slow-roll parameters for a product potential, 
$W=e^{4b(\vp)}V(\chi)$.  
In these specific models, 5 parameters are reduced
further to only 2 and 3 independent parameters for model 1 and 2 respectively.
$\epsilon_\vp$ and $\epsilon_\chi$ are directly related to $\epsilon$ 
through $\theta$.  
The three remaining parameters ($\eta_{\chi\chi}, \ \xi^2_{\chi\chi\chi}, \ \xi_b$) are shown to be also functions
 of $\epsilon$ through $\epsilon_\vp$ and $\epsilon_\chi$.} 
\label{Tab1}
\end{center}
\end{table}

\subsection{Jordan-Brans-Dicke theory}
First we apply our results to the Jordan-Brans-Dicke (JBD) theory with quadratic and quartic 
potentials.
In this theory, 
\begin{eqnarray}
U(\vp)=e^{4b(\vp)}, \quad \quad b(\vp)=-\beta\frac{\vp}{\Mp},
\label{Model1}
\end{eqnarray}
where we assume $\beta$ is a positive constant.  For this potential choice, $\ep=\eb=\eta_{\vp\vp}/2=8\beta^2$ 
and the slow-roll ends when $\epsilon^e=\ec^e+\ep^e=1$.

Two potentials are chosen for $\chi$:
\begin{itemize}
\item {\bf Model 1a} $V(\chi)=\frac{\lambda}{4}\chi^4$
\item {\bf Model 1b} $V(\chi)=\frac12m^2_\chi \chi^2$
\end{itemize}
where $\lambda$ is a dimensionless parameter and $m_\chi$ is the mass of the $\chi$ field.
The slow-roll parameters for both quartic and quadratic potentials for the $\chi$-field are shown in the first two columns of Table~\ref{Tab1}.
As seen in the table, the choice of potentials reduces the 5 independent slow-roll parameters to only two: $\epsilon_\vp$ and $\epsilon_\chi$.  
Equivalently, since $\epsilon_\vp=\epsilon\cos^2\theta$ and $\epsilon_\chi=\epsilon\sin^2\theta$, we can use the parameters $\epsilon$ and $\theta$.  
Hence, all the first and second order slow-roll parameters are proportional 
to $\epsilon$ and $\epsilon^2$ respectively 
which make the primordial spectral indices proportional
 to $\epsilon^2$ in the lowest order.

One final parameter is required, $\tan\Delta$, to describe the evolution after inflation. 
Then $\frac{1}{\epsilon^2}\frac{dn_{(\zeta,\ C,\ S)}}{d\ln k}$ are just 
a function of $\theta$ and $\Delta$. We note that $\theta$ is calculated 
at horizon crossing and depends on $\beta$ and the initial values 
of $\vp$ and $\chi$, whereas $\Delta$ represents the evolution after horizon 
crossing and depends on the late time evolution of the Universe.

\begin{figure}[!t]
%  \begin{center}
  \begin{tabular}{c c}
    \includegraphics[width=0.33\textwidth]{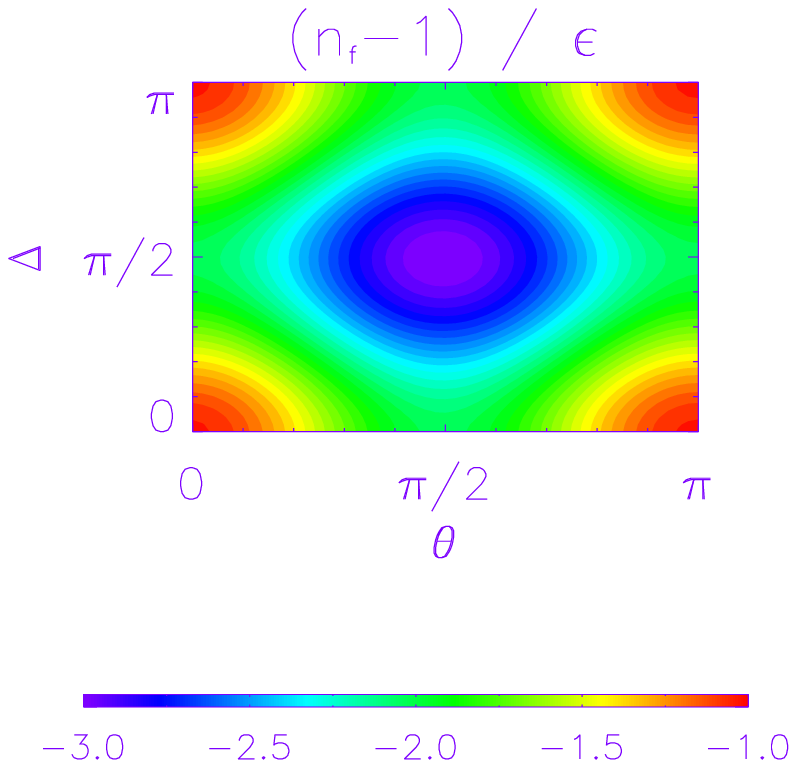} 
&
    \includegraphics[width=0.66\textwidth]{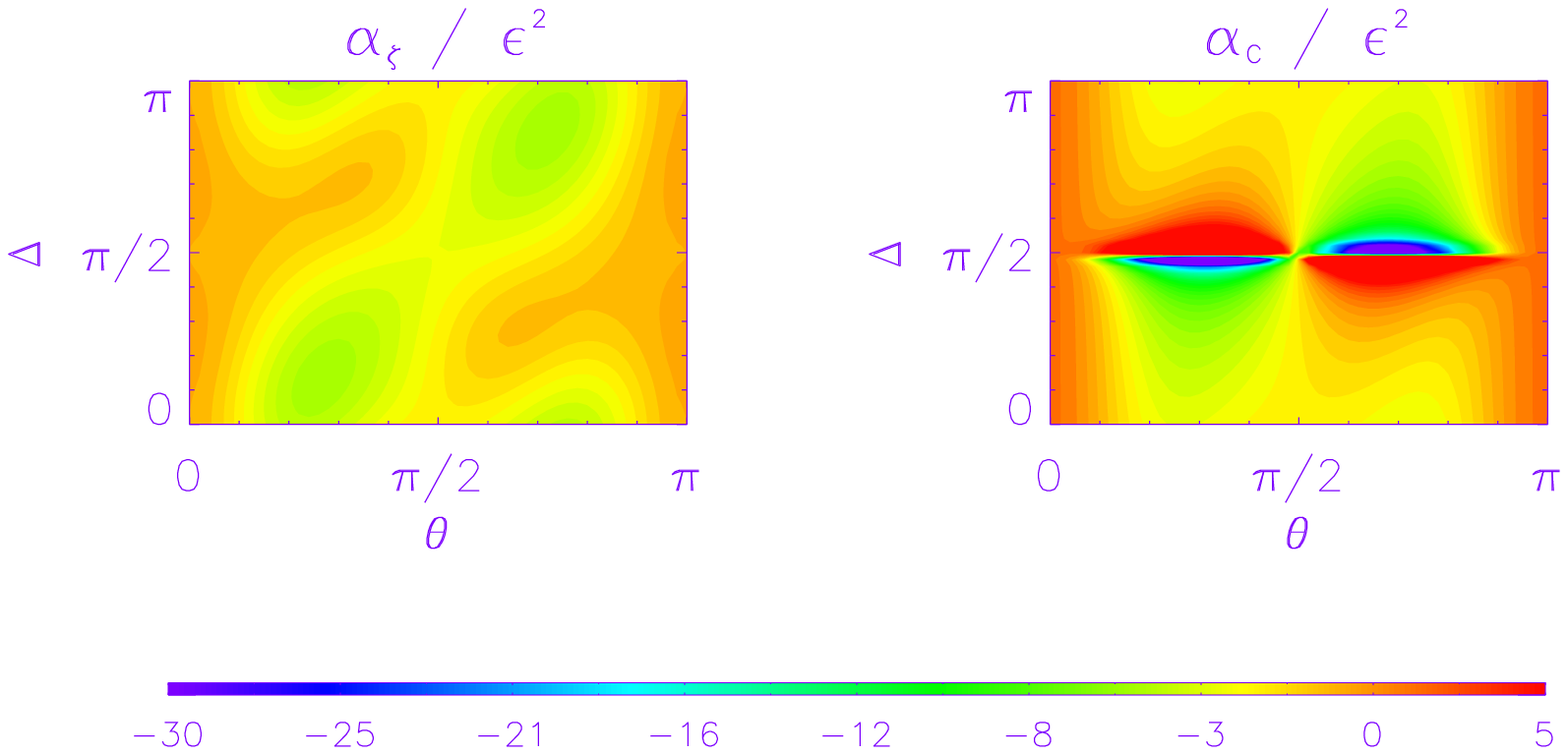}
%&
%    \includegraphics[width=0.5\textwidth]{./figs/case1b.ps} 
    \end{tabular}
%  \end{center}
  \caption{The spectral index ($(n_\zeta -1)/\epsilon$) and running of $n_\zeta$ and $n_C$ in terms of $\epsilon$ for model 1-a.}
  \label{fig:1-ab}
\end{figure}

In Figure~\ref{fig:1-ab} we show the running spectral indices $\alpha_\zeta$ and
$\alpha_C$ for Model 1a.  
The result for Model 1b is almost identical and is not shown. 
As said before, $\alpha_S$ is independent of $\Delta$, thus we plot
the $\alpha_S$ dependence on $\theta$ separately in Figure~\ref{fig:runningS}.
We plot in the range of $0<\theta <\pi$ since the potential has 
the reflection symmetry $\chi \rightarrow -\chi$ which corresponds
to the change of $\theta$ to $-\theta$.

For the correlated spectral running, $\alpha_C$, there is an divergence 
at $\Delta\rightarrow\frac{\pi}{2}$.  This divergence is not observable. 
From Eq.~(\ref{Pzeta}), it can be seen that, for $\Delta=\frac{\pi}{2}$ ($T_{\zeta S}=0$), there is no evolution in $\calP_\zeta$ and the amplitude of the correlation
spectrum is zero.
It is therefore impossible to define a spectral index or running at this point.
As $\Delta\rightarrow\frac{\pi}{2}$, $T_{\zeta S}$ is non-zero, but very small.  
In this limit, the correlation amplitude would be also small and therefore unobservable.

As an aside, it can be noted that, for the specific case of Model 1a, 
$B$ (defined in Eq.~(\ref{Beq})) is identically zero, 
due to a cancelling of the slow-roll parameters.
This means that from Eq.~(\ref{evol}) the curvature perturbation remains 
constant after horizon crossing ($\dot{\zeta}=0$) 
for this JBD model with quartic potential.

It is possible to consider the case $V(\chi)= \lambda \chi^n$. 
In this JBD theory, and using the fact that at the end of inflation $\epsilon^e=\epe+\ece=1$, we can analytically solve $N$ in the slow-roll limit,
\dis{
N=\frac{1}{8\beta^2}\ln \left[1+\frac{4\beta^2}{n\Mp^2}e^{-2\beta\vp_*/\Mp}\chi_*^2 \right]+\frac{1}{8\beta^2}\ln \left[\frac{1-8\beta^2}{1+(2n-8)\beta^2} \right].
\label{Nsolve}}
We find that $\beta<0.05$ is required to obtain enough e-foldings for $\vp^*, \chi^* <50\Mp$.  
From this fact, $8\beta^2 N \lesssim 1$ and the limit of Eq.~(\ref{Nsolve}) gives $\epsilon^*\simeq \frac{n}{4N}< 0.004 n$, where we have taken $N\approx 60$.
In this limit of $\beta$, we can approximate Eq.~(\ref{Nsolve}) further:
\begin{equation}
N\approx\frac{\chi_*^2}{2n}-\frac{n}{4} -\beta \frac{\vp_*\chi_*^2}{n}.
\label{approxNmod1}
\end{equation}

For Models 1a and 1b, we see from Figure~\ref{fig:1-ab} that 
\dis{
-10 (\epsilon^*)^2 \lesssim \alpha_\zeta \lesssim 0.
}
Hence $|\alpha_\zeta|\lesssim 1.6 n^2 \times10^{-4}$
which is quite compatible with observations~\cite{Spergel:2006hy}.

From the slow-roll equations of motion, Eq.~(\ref{slow-roll}), we find the solution of $\vp$
\dis{
\vp = \vp_*+\Mp {\rm sign}\left(\frac{U_{\vp}}{U}\right)\sqrt{2\ep}N,
}
which gives $b_e-b_*=-4\beta^2N$.
Using this we can express Eq.~(\ref{fNL4}) at the end of inflation 
(we take $\epsilon^e=\epsilon^e_\chi+\epsilon^e_\vp=1$) in terms of slow-roll parameters at horizon crossing.
In the limit $8\beta^2N\ll 1$, we find 
$\alpha\approx 1+N$, $u\approx\epsilon^e_\vp$ and $v\approx1-\epsilon^e_\vp$.
This leads to the simplified form of $f^{(4)}_{NL}$,
\dis{
-\frac65 f^{(4)}_{NL}\simeq 2\ecstar-\eta_{\chi\chi}^*+2\ecstar\epstar(N+1)-\epstar.
\label{fNLmodel1}
} 
It can be noted that in this small $\beta$ limit, $\epstar<\ecstar$ and the last 
two terms in this equation are negligible compared to the first two terms.  

Using Eq.~(\ref{approxNmod1}), $f_{\rm NL}^{(4)}$ can be written in terms of the required e-folding number in the small $\beta$ limit (required for enough inflation)
\begin{equation}
-\frac65 f^{(4)}_{NL} \simeq \frac{1}{2N} + \frac{\beta}{N} \frac{\vp_*}{\Mp}.
\end{equation}
It is clear that, in order to achieve enough inflation, the parameters are required to be small and hence $-\frac65 f^{(4)}_{NL}\approx \frac{1}{2N}$, which is unobservable.

\subsection{Brans-Dicke Type Models with Quadratic Exponent}
In order to observe the effect of higher powers in $b(\vp)$, we consider a quadratic function, such that $\xi_b\neq0$.
To this end, we consider a Brans-Dicke type model with  
\begin{eqnarray}
U(\vp)=e^{4b(\vp)}, \quad \quad b(\vp)=-\beta\vp^2/\Mp^2.
\label{Model2}
\end{eqnarray}
With this choice, $\ep=32\beta^2\vp^2/\Mp^2$ and $\eta_{\vp\vp}=-8\beta+64\beta^2\vp^2/\Mp^2$.

Once again, two example potentials are chosen for $\chi$:
\begin{itemize}
\item {\bf Model 2a} $V(\chi)=\frac{\lambda}{4}\chi^4$
\item {\bf Model 2b} $V(\chi)=\frac12m^2_\chi \chi^2$
\end{itemize}
where $\lambda$ and $m_\chi$ have the same definitions as before.

\begin{figure}[!t]
%  \begin{center}
  \begin{tabular}{c c}
    \includegraphics[width=0.33\textwidth]{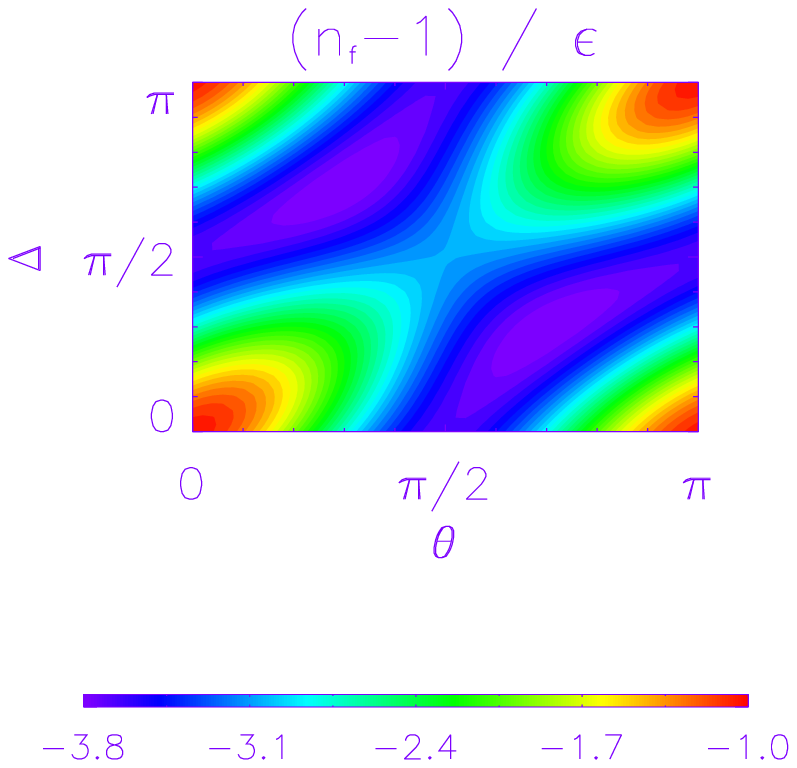} 
&
    \includegraphics[width=0.66\textwidth]{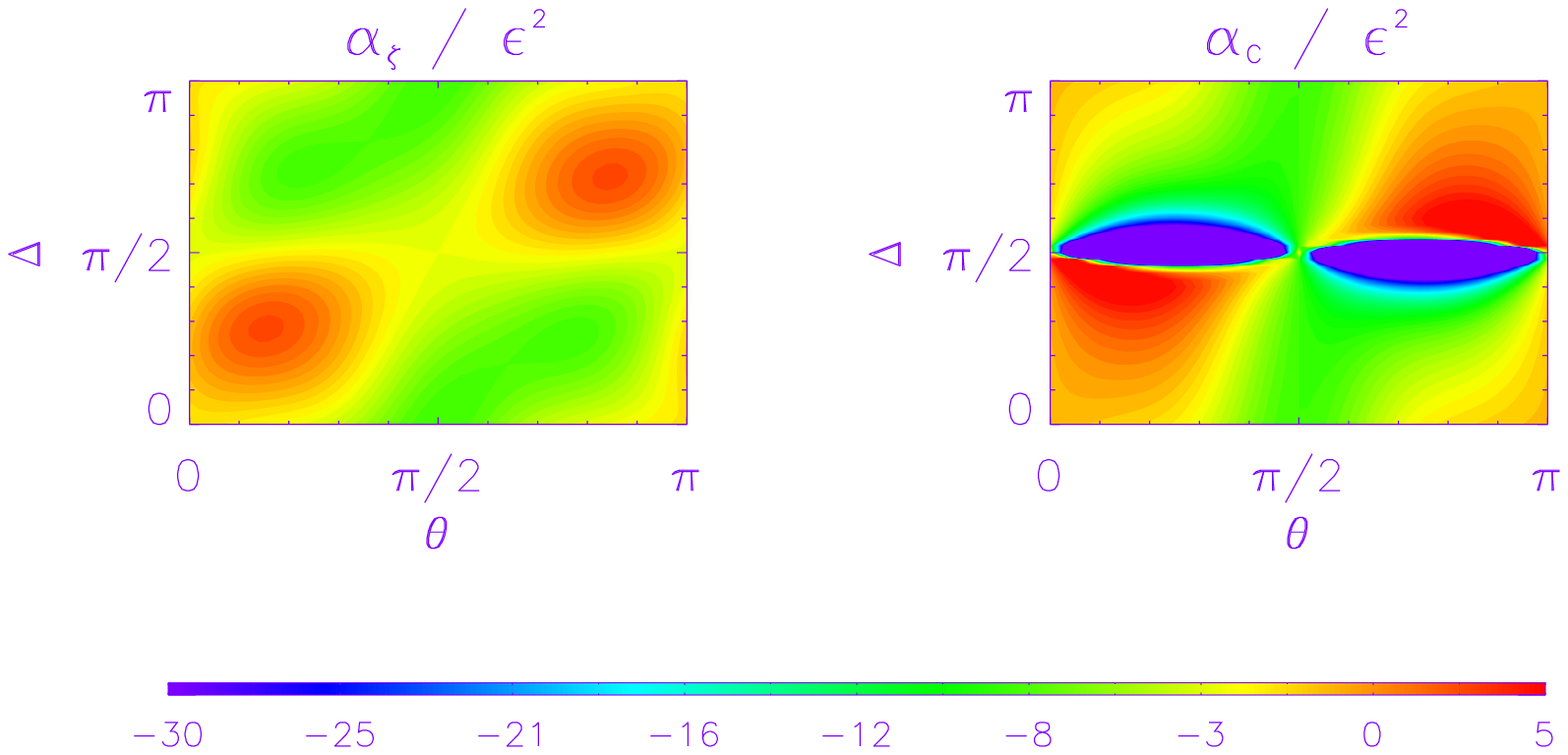} 
%&
%    \includegraphics[width=0.5\textwidth]{./figs/case2b.ps} 
    \end{tabular}
%  \end{center}
  \caption{Spectral index ($n_\zeta -1$) and running of $n_\zeta$ and $n_C$ for model 2-a. We used $\beta=0.1\epsilon$ here.}
  \label{fig:2ab}
\end{figure}

\begin{figure}[!t]
  \begin{center}
    \includegraphics[width=0.4\textwidth]{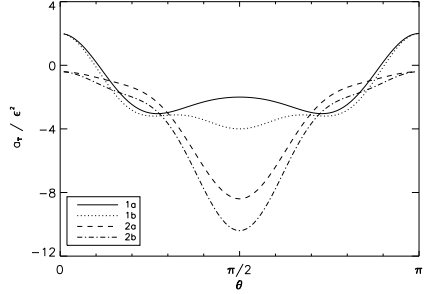} 
  \end{center}
  \caption{Running of $n_S$ for Models 1 and 2. $\alpha_S$ is independent of $\Delta$.}
  \label{fig:runningS}
\end{figure}

The slow-roll parameters for these potentials are shown in the last two columns of Table~\ref{Tab1}.  Once again, most of the slow-roll parameters can be written in terms of the single parameter, $\epsilon^2$.  The exception is $\xi_b$, which is related to the coefficient of the coupling term, $\beta$.  In order to analyse the expected levels of running from these cases, it is necessary to relate $\beta$ to $\epsilon$ and we take the assumption that $\beta=0.1\epsilon$. 
For Model 2a, the numeric calculations of the spectra for this special case (in relation to $\epsilon$, $\theta$ and $\Delta$) are shown in Figures~\ref{fig:2ab} and \ref{fig:runningS}.  Model 2b is almost identical to Model 2a and the numerical results are not shown. 

Models 2a and 2b have a reflection symmetry about $\vp$ 
($\theta\rightarrow -\theta$) and/or $\chi$ ($\theta\rightarrow \pi-\theta $).
However since the sign of $\tan\Delta$ is determined by the sign of 
$\alpha$ in Eq.~(\ref{evolution-of-fluctuations}) and
 $\alpha$ is proportional to $\sin 2\theta$ for the potential in 
Eq.~(\ref{app1}), The reflection of $\vp$ or $\chi$
changes also the sign of $\tan\Delta$ before the end of inflation.
For this reason, we see the symmetry in Figure~\ref{fig:2ab} 
around the point ($\theta=\pi/2,\Delta=\pi/2$).
However after inflation, $\Delta$ is no longer dependent on the fields 
$\vp$ and $\chi$, thus this symmetry is not necessarily valid any more.

In this model, the solutions for $\vp$ and $\epsilon_\vp^e$ are given by
\dis{
\vp_e=\vp_*e^{8\beta N}, \qquad  
\epsilon_\vp^e=32\beta^2\vp_e^2/\Mp^2 = \epsilon_\vp^* e^{16\beta N}.
\label{model2solphi}
}
%and
%\dis{
%\chi_e^2=\chi_*^2-\frac{n}{8\beta}e^{2\beta_*^2}\left(e^{16\beta N}-1 \right).
%} 
Due to the quadratic coupling, the potential is steeper than that of traditional Brans-Dicke models, so that a much smaller
$\beta$ is required to give enough e-folding number.
In this case, if we assume $16\beta N \ll 1$,
then we obtain
\dis{
N\simeq\frac{1}{16\beta}\ln\left[1+\frac{8\beta}{n}e^{-2\beta\vp_*^2}
\chi_*^2\right]-\frac{1}{16\beta}\ln\left[1+\frac{4\beta n^2}{1-32\beta^2\vp_*^2} \right].
\label{Nmod2}
}
In order to obtain enough inflation, for $\vp_*,\chi_*<50\Mp$, then we require $\beta<0.0005$.  In this very small limit, we can approximate Eq.~(\ref{Nmod2}):
\begin{equation}
N\approx\frac{\chi_*^2}{2n}-\frac{n^2}{4} +\beta
\left(
\frac{n^4}{2}-\frac{\vp_*^2\chi_*^2}{n} - \frac{2\chi_*^4}{n^2}
\right).
\label{approxNmod2}
\end{equation}

As in Model 1a, it is possible to estimate the non-Gaussianity for small $\beta$.
Again, we take $\epsilon^e=1$, so that $\alpha$, $u$ and $v$ are given as before.
In the limit $16\beta N\ll1$, we find
\dis{
-\frac65 f^{(4)}_{NL}\simeq 2\ecstar-\eta_{\chi\chi}^*+2\ecstar\epstar(N+1)-\epstar-\frac12\left(\ecstar\right)^2\epstar\xi_b(N+1)^3.
\label{fNLmodel2}
}
The final term is directly due to $b_{\vp\vp}\neq0$.  Using the approximation in Eq.~(\ref{approxNmod2}), we find
\begin{equation}
-\frac65 f^{(4)}_{NL} \simeq \frac{1}{2N} + \frac{\beta}{N} \frac{\vp_*^2}{\Mp^2}.
\end{equation}
As in Model 1, in order to achieve enough inflation, the parameters are required to be small and hence $-\frac65 f^{(4)}_{NL}\approx \frac{1}{2N}$.

\section{Examples of Scalar-Tensor Theories with Sum Potentials}
\label{SecTheory2}
The second models we consider are those with sum potentials.  If we define the potential
\dis{
W(\vp,\chi)=U(\vp)+V(\chi)=\frac12m_\vp^2\vp^2+\frac12m^2_\chi\chi^2,
}
then the slow-roll parameters simplify greatly and are given by
\dis{
\epsilon_\vp=\frac{\Mp^2}{2}\left(\frac{m^2_\vp \vp}{W}\right)^2, \qquad 
\epsilon_\chi=\frac{\Mp^2}{2}\left(\frac{m^2_\chi \chi}{W}\right)^2e^{-2b},
}
\dis{
\eta_{\vp\vp}=\Mp^2\frac{m^2_\vp}{W},\qquad 
\eta_{\chi\chi}=\Mp^2\frac{m^2_\chi}{W}e^{-2b},\qquad \eta_{\vp\chi}=0,
}
\dis{
\xi^2_{\vp\vp\vp}=\xi^2_{\vp\vp\chi\chi}=\xi^2_{\vp\chi\chi\chi}=\xi^2_{\chi\chi\chi}=0.}

It is no longer possible to directly relate the parameters to $\epsilon$ and $\theta$.
However, if we represent the ratio of masses by
\dis{
r=\frac{m_\chi^2}{m_\vp^2},}
then the slow-roll parameters can be re-written in terms of $\epsilon$, $\beta$ and $r e^{-2b}$ only:
\dis{
\eta_{\vp\vp}=\epsilon\left(\cos^2(\theta)+\frac{\sin^2(\theta)}{r e^{-2b}}\right), \qquad
\eta_{\chi\chi}=\epsilon\left(\sin^2(\theta)+\cos^2(\theta)r e^{-2b}\right), \qquad
}
\dis{
\epsilon_b=64\frac{\beta^2}{\epsilon}\frac{\cos^2(\theta)}{\left(\cos^2(\theta)+\frac{\sin^2(\theta)}{r e^{-2b}} \right)}.}

\begin{figure}[!t]
%  \begin{center}
  \begin{tabular}{c c}
    \includegraphics[width=0.33\textwidth]{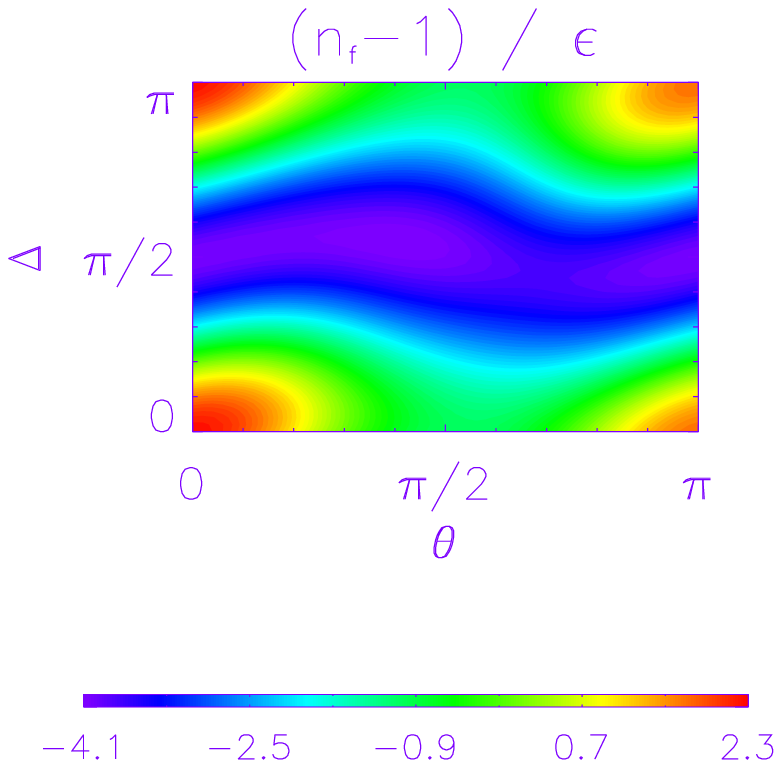} 
&
    \includegraphics[width=0.66\textwidth]{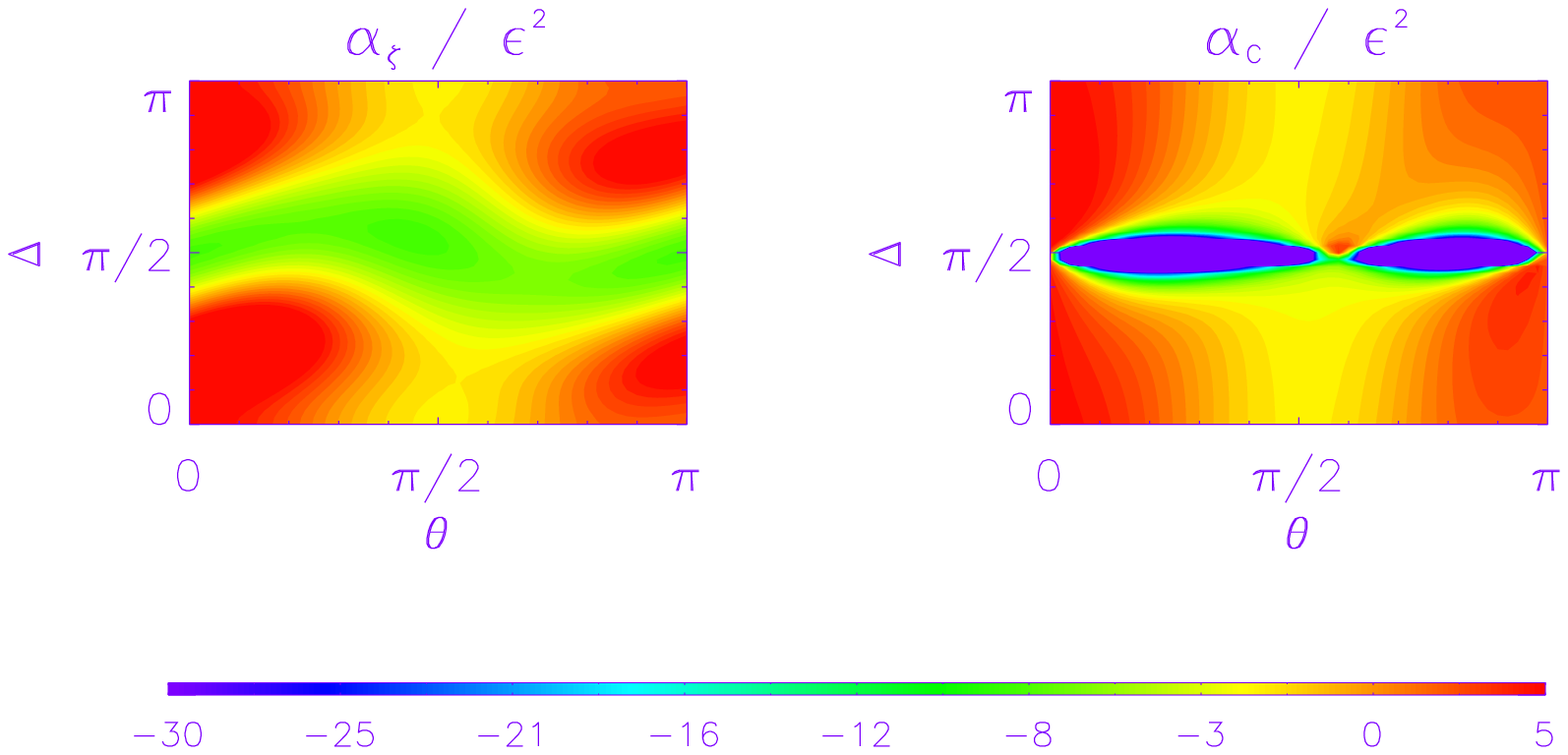}
%&
%    \includegraphics[width=0.5\textwidth]{./figs/case4.ps} 
    \end{tabular}
%  \end{center}
  \caption{Spectral index ($n_\zeta -1$) and running of $n_\zeta$ and $n_C$ in terms of $\epsilon$ for model 3. Values of $\beta=0.1\sqrt{\epsilon}$ and $r e^{-2b}=2$ have been assumed.}
  \label{fig:34}
\end{figure}

\begin{figure}[!t]
  \begin{center}
    \includegraphics[width=0.4\textwidth]{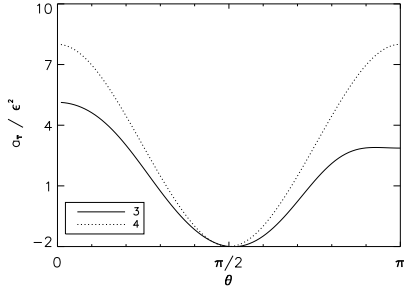} 
  \end{center}
  \caption{Running of $n_S$ for Models 3 and 4. $\alpha_S$ is independent of $\Delta$.}
  \label{fig:runningS34}
\end{figure}

For this sum potential, the remaining choice is in the function of $b(\vp)$ and we consider the two obvious cases:
\begin{itemize}
\item {\bf Model 3} $b(\vp)=-\frac{\beta \vp}{\Mp}$ 
\item {\bf Model 4} $b(\vp)=-\frac{\beta \vp^2}{\Mp^2}$
\end{itemize}

We can fix $\beta$ and the mass ratio, in order to estimate the spectral index and running.  We take $r e^{-2b}=2$ and, for Models 3 and 4 respectively, we assume $\beta=0.1\sqrt{\epsilon}$ and $\beta=0.1\epsilon$.
The results of the spectrum for Model 3 are shown in Figures~\ref{fig:34} and \ref{fig:runningS34}.
The results for Model 4 are very similar to those of Model 3 and are therefore not shown.

\section{Numerical Analysis of Non-Gaussianity}
\label{secNumerical}
In order to verify our analytical results of the previous section, we numerically solved the background equations of motion, Eq.~(\ref{background}) and the Friedmann equation, Eq.~(\ref{Friedmann}), not assuming slow-roll.  
Multiple trajectories were considered, with initial values, $\vp_i$ and $\chi_i$, given by a $\vp$-$\chi$ grid and $\dot\vp_i=\dot\chi_i=0$.  
The end of inflation is taken to be the time at which the slow-roll parameter $\epsilon=\ep+\ec$ becomes unity and the total number of e-foldings until this point is given by $N_f$.  

At horizon crossing, the fields have reached values of $\vp_*$ and $\chi_*$ and the remaining number of e-folds is denoted by $N$.  
The calculation of non-Gaussianity requires the gradient of the number of e-foldings between horizon crossing and the end of inflation, $N$ with respect to each of the fields at crossing, $\vp_*$ and $\chi_*$.

It is useful to notice, quite trivially, that if for a given trajectory, $\vp_i=\vp_*$ and $\chi_i=\chi_*$, then $N_f=N$.
It is therefore possible to calculate $N,_i$, $N,_{ij}$ etc by differentiating the grid of $N_f$ with respect to $\vp_i$ and $\chi_i$.

\begin{figure}[!t]
  \begin{tabular}{c c}
    \includegraphics[width=0.5\textwidth]{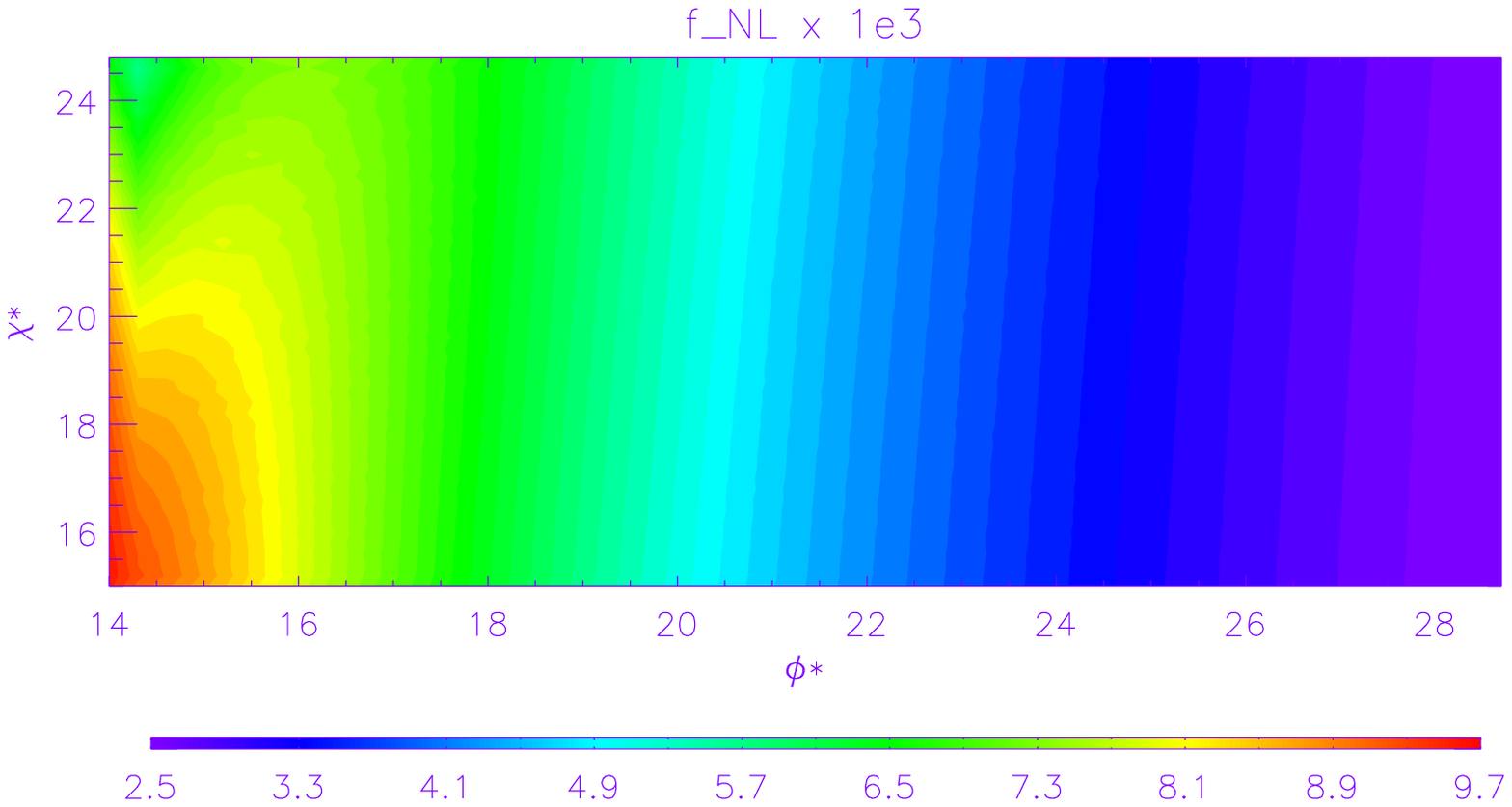}
&
    \includegraphics[width=0.5\textwidth]{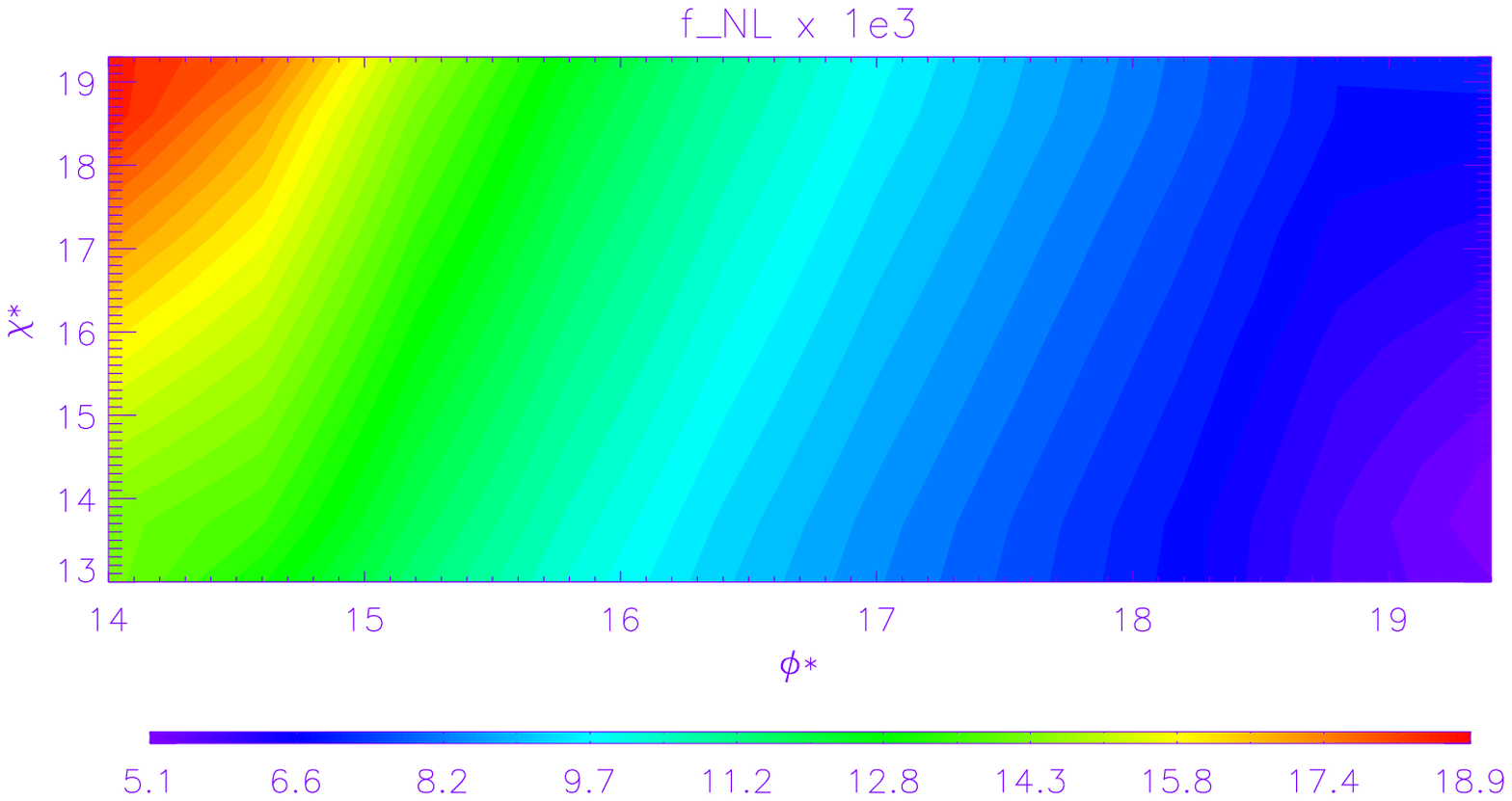}
    \end{tabular}
  \caption{Level of non-Gaussianity for Models 3 (left) and 4 (right) calculated from the second term of Eq.~(\ref{fNL}). We have assumed $m_\vp=m_\chi=5\times 10^{-5} \Mp$. $\beta=0.1$.}
  \label{fig:f_nl}
\end{figure}

Each of the cases considered above (Models 1-4) were modelled numerically.
The analytical slow-roll approximations given in Eq.~(\ref{fNLmodel1}) and (\ref{fNLmodel2}) were seen to agree within $5\%$.
For Models 3 and 4, no analytical approximation was found.  For these sum potentials, the numerical level of non-Gaussianity was calculated from the second term in Eq.~(\ref{fNL}) and the results are shown in Figure~\ref{fig:f_nl}.

In previous work, \cite{VernizziWands}, for theories with {\it canonical} kinetic terms and a sum potential, $f_{\rm NL}\approx1/N$.  
We find numerically that the $f_{\rm NL}$ is of that order of magnitude, even with non-canonical couplings.  Therefore we conclude that $\beta$ has little impact on the level of nonlinearity.

\section{Conclusion}
\label{SecConc}
The focus of this paper was slow-roll inflation in theories with two scalar fields, 
coupled through a potential as well as their kinetic terms. We have derived the general 
formulae for the running of the spectral indices for adiabatic and entropic perturbations 
as well as for the cross-correlation spectrum. In addition, using the $\delta N$-formalism and 
specialising to the product potential ($W(\varphi,\chi)=U(\varphi)V(\chi)$) and the sum potential 
($W(\varphi,\chi)=U(\varphi)+V(\chi)$) we have derived the general expressions for the 
nonlinearity parameter $f_{\rm NL}$ during inflation. 

We have calculated the spectral indices and runnings for several example models. The results are within present observational ranges.

One of the specific examples we have considered is the Jordan-Brans-Dicke theory with quadratic and 
quartic potentials. 
In this case, the slow-roll parameter for the field $\varphi$ is 
given by the (constant) coupling parameter $\beta$, defined in Eq.~(\ref{Model1}). The slow-roll 
condition for the field $\varphi$ implies that $\beta$ itself has to be small. In this case, 
an analytical formula for $f_{\rm NL}$ during inflation is given by Eq.~(\ref{fNLmodel1}) and is of order $1/N$, where $N$ is the total number of e-foldings.
Assuming $\beta$ to be positive, the effect of the coupling is to {\it enhance} $f_{\rm NL}$ by an additional term $\frac{\beta\varphi}{N\Mp}$. 
When the coupling is quadratic in the $\vp$-field (as in Eq.~(\ref{Model2})), the nonlinearity is {\it enhanced} by a term $\frac{\beta\varphi^2}{N\Mp^2}$.  Neither of these terms are significant and in the slow-roll limit, $f_{\rm NL}$ is not much enhanced by the coupling.

We also considered models with a sum of two quadratic potentials.  Due to the complexity of the analytic formula for $f_{\rm NL}$, it is difficult to find an approximate equation.  
We therefore provide numerical results for simple models and showed that $f_{\rm NL}\ll1$ and is unobservable for the slow-roll limit.

To conclude, in order to distinguish between inflationary models, future observations will search for 
deviations from standard one-field inflation. In order to do so one has to go beyond the spectral index as 
an observable. 
Future experiments, such as PLANCK, will tighten the constraints on the spectral running and nonlinearity.  
The results of this paper can be used to calculate these parameters in other multi-field models with non-canonical kinetic terms, in order to
constrain the parameter ranges.

\acknowledgments
CvdB, K-YC and LH acknowledge support from PPARC. 

\appendix
\section{Slow-Roll Parameters}
\label{App1}
The slow-roll parameters we shall use are first-order
\dis{
\epsilon=-\frac{\dot{H}}{H^2}=\frac{1}{2\Mp^2}\frac{\dot{\sigma}}{H^2}=\frac{\Mp^2}{2}\left(\frac{V_\sigma}{V}\right)^2=\epsilon_\vp+\epsilon_\chi,
}
\dis{
\epsilon_\vp
=\frac{\Mp^2}{2}\left(\frac{V_\vp}{V}\right)^2=\epsilon \cos^2\theta,\qquad
\epsilon_\chi
=\frac{\Mp^2}{2}\left(\frac{V_\chi}{V}\right)^2e^{-2b}=\epsilon \sin^2\theta,
}

\dis{
\epsilon_b=8\Mp^2b_\vp^2,
}

\dis{
\etagg&\equiv\eta_{\vp\vp}\cos^2\theta+\eta_{\vp\chi}\sin2\theta+\eta_{\chi\chi}\sin^2\theta,\\
\etags&\equiv(\eta_{\chi\chi}-\eta_{\vp\vp})\sin\theta\cos\theta + \eta_{\vp\chi}(\cos^2\theta-\sin^2\theta),\\
\etass&\equiv \eta_{\vp\vp}\sin^2\theta-\eta_{\vp\chi}\sin2\theta+\eta_{\chi\chi}\cos^2\theta,}
where
\dis{
\eta_{\vp\vp}=\Mp^2\frac{V_{\vp\vp}}{V},\qquad\eta_{\vp\chi}=\Mp^2\frac{V_{\vp\chi}}{V}e^{-b},\qquad \eta_{\chi\chi}=\Mp^2\frac{V_{\chi\chi}}{V}e^{-2b},
}
and second-order
\dis{
\xisqggg&\equiv \xi^2_{\vp\vp\vp}\cos^2\theta+\xi^2_{\vp\vp\chi\chi}3\cos^2\theta
+\xi^2_{\vp\chi\chi\vp}3\sin^2\theta+\xi^2_{\chi\chi\chi}\sin^2\theta,\\
\xisqggs&\equiv (\xi^2_{\vp\vp\vp}-\xi^2_{\chi\chi\chi})\sin\theta\cos\theta
+\xi^2_{\vp\vp\chi\vp}(2\sin^2\theta-\cos^2\theta)
+\xi^2_{\vp\chi\chi\chi}(\sin^2\theta-2\cos^2\theta),\\
\xisqgss&\equiv \xi^2_{\vp\vp\vp}\sin^2\theta+\xi^2_{\vp\vp\chi\chi}(2\cos^2\theta-\sin^2\theta)+\xi^2_{\vp\chi\chi\vp}(-\cos^2\theta+2\sin^2\theta)+\xi^2_{\chi\chi\chi}\cos^2\theta,
}
where
\dis{
\xi^2_{\vp\vp\vp}=\Mp^4\frac{V_{\vp\vp\vp} V_\vp}{V^2},\qquad
&\xi^2_{\chi\chi\chi}=\Mp^4\frac{V_{\chi\chi\chi} V_\chi e^{-4b}}{V^2},\qquad \\
\xi^2_{\vp\vp\chi\chi}=\Mp^4\frac{V_{\vp\vp\chi} V_\chi e^{-2b}}{V^2},\qquad
&\xi^2_{\vp\vp\chi\vp}=\Mp^4\frac{V_{\vp\vp\chi} V_\phi e^{-b}}{V^2} = \xi^2_{\vp\vp\chi\chi}\cot\theta,\qquad \\
\xi^2_{\vp\chi\chi\chi}=\Mp^4\frac{V_{\vp\chi\chi} V_\chi e^{-3b}}{V^2},\qquad
&\xi^2_{\vp\chi\chi\vp}=\Mp^4\frac{V_{\vp\chi\chi} V_\vp e^{-2b}}{V^2} = \xi^2_{\vp\chi\chi\chi}\cot\theta.
}

Here we note that $\xi^2_{sss}$ does not appear in the running of spectral indexes.

\dis{
\xi_b=8\Mp^2b_{\vp\vp}.
}

\section{Time Derivative of First Order Slow-Roll Parameters}
\dis{
\frac{1}{H}\dot{\epsilon}=4\epsilon^2-2\epsilon\etagg+\frac12\rtebe\sin^2\theta\cos\theta\epsilon
}
\dis{ 
\frac{1}{H}\dot{\epsilon_b}=-\rtebe\cos\theta\xi_b
}
\dis{
-\frac{\ddot{H}}{H^3}=-2\epsilon^2+\frac{1}{H}\dot{\epsilon}
}
\dis{
\frac{1}{H}\dot{\eta}_{\sigma\sigma}=&2\epsilon\etagg-2\etags^2-\xisqggg\\
&+(\etagg\sin\theta+2\etags\cos\theta)\sin\theta\cos\theta\rtebe\\
\frac{1}{H}\dot{\eta}_{\sigma s}=&2\epsilon\etags+\etags(\etagg-\etass)
-\xisqggs\\
&+(\frac12\etags\cos\theta+\etass\cos^2\theta\sin\theta)\rtebe\\
\frac{1}{H}\dot{\eta}_{ss}=&2\epsilon\etass+2\etags^2-\xisqgss+\etass\cos^3\theta\rtebe
}

\dis{
\frac{\partial}{H_*\partial t_*}\ln(1+T^2_{\zeta S})&=2\sinD\cosD(-\alpha_*-\delta_*T_{\zeta S})\\
\frac{\partial}{H_*\partial t_*}\ln T_{S S}&=-\delta_*\\
\frac{\partial}{H_*\partial t_*}\ln T_{\zeta S}&=\frac{-\alpha_*-\delta_*T_{\zeta S}}{T_{\zeta S}}
}

\section{Calculation of $-\frac65f_{\rm NL}^{(3)}$ for Non-Canonical Terms}
\label{AppC}
The first term of Eq.~(\ref{fNL}), $f_{\rm NL}^{(3)}$, can be written by
\dis{
-\frac65f_{\rm NL}^{(3)}=\frac{\calA^{IJK}N_{,I} N_{,J} N_{,K}}
{( N_{,I} N_{,J}\calG^{IJ})^2\sum_i k_i^3 },\label{fnl3} 
}
where  summations over $I,J,K$ are implied and  $\calA^{IJK}$ is given 
by~\cite{Seery05}
\dis{
\calA^{IJK}(k_1,k_2,k_3)=\frac{1}{\Mp^2}\sum_{perms}\frac{\dot{\vp}^I}{4H}
\calG^{JK}M_{123},
}
with momentum dependent $\calM_{123}=\calM(k_1,k_2,k_3)$.
We can perform the sum and permutations in the numerator of 
Eq.~(\ref{fnl3}):
\dis{
  \sum_{IJK}\calA^{IJK}N_{,I} N_{,J} N_{,K}=\sum_{IJK}N_{,I} N_{,J} N_{,K}\frac{1}{\Mp^2}\sum_{perms}\frac{\dot{\vp}^I}{4H}
\calG^{JK}M_{123}.
}
Following from this, the numerator in Eq.~(\ref{fnl3})
can be reduced to
\dis{
-\frac{1}{4\Mp^2}\left(\sum_{IJ}N_{,I}N_{,J}\calG^{IJ}\right)\sum_{perms}\calM_{123}.
}
where we have used  $\sum_IN_{,I}\dot{\vp}_I=-H$.
We define
\dis{
f=f(k_1,k_2,k_3)\equiv -1-\frac{\sum_{perms}\calM_{123}}{2\sum_i k_i^3}.
}
Finally, if we consider the power spectra of curvature perturbations
and gravitational waves,
\dis{
\calP_\zeta=\calP_*\sum_{IJ}N_{,I}N_{,J}\calG^{IJ},\qquad \calP_*\equiv\frac{H_*^2}{4\pi^2},\qquad
\calP_g=\frac{8\calP_*}{\Mp^2},
}
we obtain\dis{
 -\frac65f_{\rm NL}^{(3)}=\frac{r}{16}(1+f),
}
where $r$ is scalar-to-tensor ratio, $r\equiv\calP_g/\calP_\zeta $.

\end{document}